\documentclass[%
 reprint,
 amsmath,amssymb,
 aps,
]{revtex4-2}

\usepackage{graphicx}
\usepackage{dcolumn}
\usepackage{bm}
\usepackage{color}
\usepackage[normalem]{ulem} 

\def\Vec#1{{\bf #1}}
\def\GVec#1{\boldsymbol #1}

\begin{document}

\preprint{APS/123-QED}

\title{Resonant interaction in chiral, Eshelby-twisted van der Waals atomic layers}

\author{Pilkyung Moon}%
 \email{Corresponding author: pilkyung.moon@nyu.edu}
\affiliation{Arts and Sciences, NYU Shanghai, 1555 Century Avenue, Shanghai 200122, China}
\affiliation{NYU-ECNU Institute of Physics at NYU Shanghai, 3663 Zhongshan Road North, Shanghai 200062, China}
\affiliation{Department of Physics, New York University, New York 10003, USA}

\date{\today}

\begin{abstract}
We study the electronic structures of
chiral, Eshelby-twisted van der Waals atomic layers
with a particular focus on a chiral twisted graphite (CTG),
a graphene stack
with a constant twist angle $\theta$ between successive layers.
We show that each CTG can host infinitely many resonant states
which arise from the interaction between the degenerate monolayer states
of the constituent layers.
Each resonant state has a screw rotational symmetry,
and may have
a smaller reduced Brillouin zone than other non-resonant states in the same structure.
And each CTG can have
the resonant states with up to four different screw symmetries.
We derive the energies and wave functions of the resonant states
in a universal form of a one-dimensional chain 
regardless of $\theta$,
and show that these states exhibit a clear optical selection rule
for circularly polarized light.
Finally, 
we discuss the uniqueness and existence of the exact center of the lattice
and the self-similarity
of the wave amplitudes of the resonant states.

\end{abstract}

\maketitle


\section{Introduction}

\begin{figure*}[ht]
\centering
\includegraphics[width=0.9\linewidth]{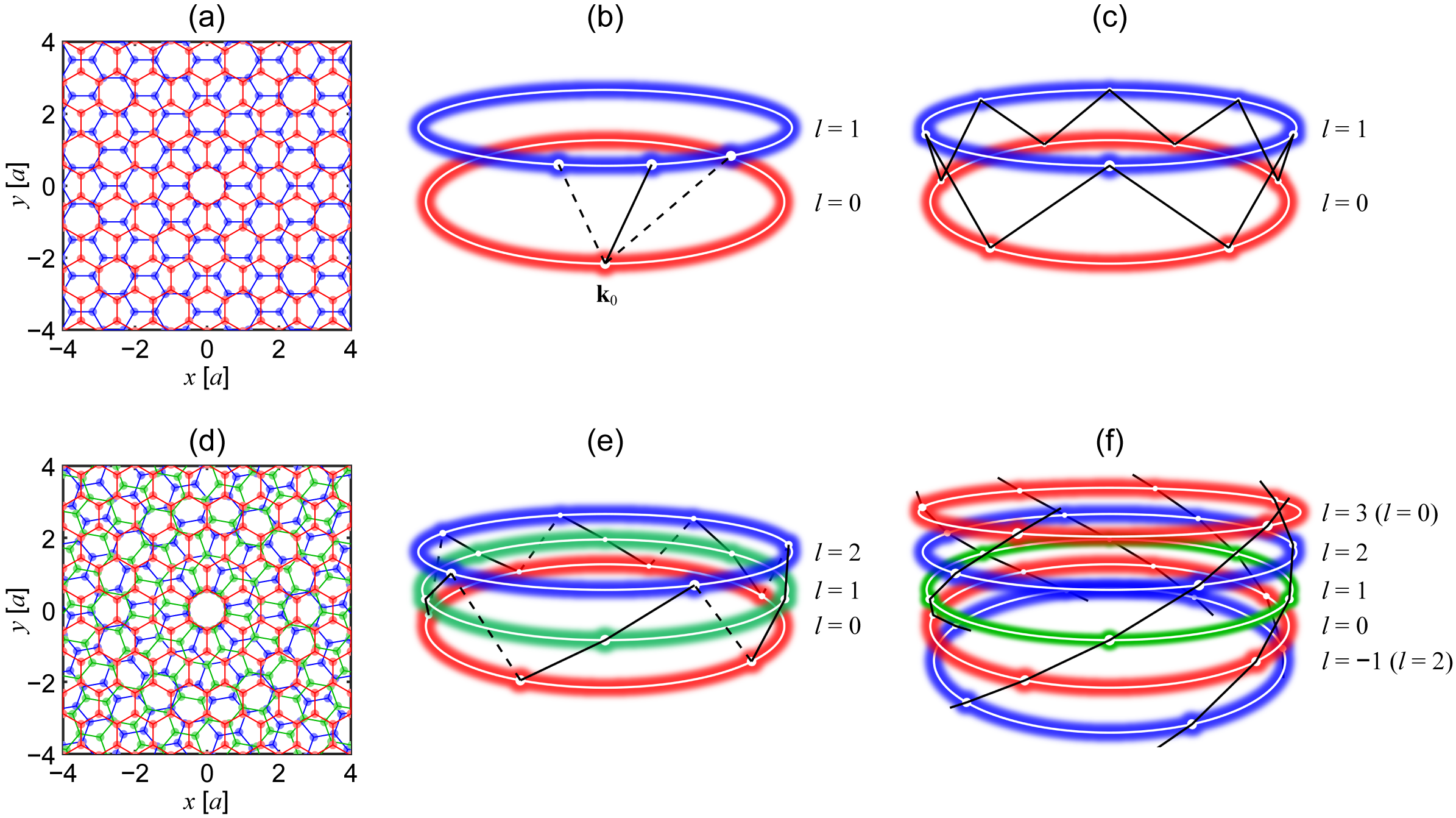}
\caption{
(a) Real-space lattice structures of quasicrystalline twisted bilayer graphene,
i.e., a bilayer graphene stacked at a twist angle $\theta=30^\circ$.
The red and blue hexagons represent the unit cells of the lower ($l=0$)
and the upper ($l=1$) layers, respectively.
(b) An illustration of the interlayer interaction at general wave vector $\Vec{k}_0$
in typical twisted bilayer systems.
The red and blue circles represent the schematic Brillouin zones of the lower and the upper layers, respectively,
with each point representing a Bloch state with a distinct Bloch wave vector.
(c) A plot similar to (b) for the quasicrystalline resonant interaction
in a quasicrystalline twisted bilayer graphene
which couples the 12 degenerate monolayer Bloch states.
(d) A plot similar to (a) for a twisted trilayer graphene
where each layer is rotated with respect to the lower layer by $\theta=20^\circ$;
the red, green, blue hexagons represent the unit cells of
the bottom ($l=0$), the middle ($l=1$),
and the top ($l=2$) layers, respectively.
(e) A resonant interaction, analogous to (c) in a bilayer system, in a trilayer system. 
The red, green, blue circles represent the schematic Brillouin zones of the bottom, the middle, and the top layers, respectively.
Although the resonant interaction couples the 18 degenerate monolayer Bloch states (points),
the interaction between the distance layers (dashed lines) is
much weaker than that between the adjacent layers (solid lines).
(f) A natural generalization of the resonant interaction in a trilayer system (c)
to an infinitely stacked CTG with $\theta=20^\circ$;
showing a portion of the entire system.
Both the lattice configurations and the Brillouin zones of the layers
with the same color are identical to each other.
The magnitudes of all the interactions represented by the solid lines are identical.
For the actual wave vectors in (e) and (f),
see Sec.~\ref{sec:reson_cond} and Figs.~\ref{Figure_04}(f)-(j).
}
\label{Figure_01}
\end{figure*}

When two atomic lattices are overlapped,
one on top of the other 
in an incommensurate configuration,
the interlayer interaction
creates an extra order along the in-plane direction
in the form of a moir\'{e} interference pattern \cite{geim2013van}.
If the two lattices have a hexagonal symmetry
[Fig.~\ref{Figure_01}(a)],
then the moir\'{e} pattern also has hexagonal symmetry
with the three dominant wave vector components
coupling one monolayer Bloch state $|\Vec{k}_0\rangle$ in either layer
to three monolayer states $|\tilde{\Vec{k}}_i\rangle$ ($i=1,2,3$) in the other layer [Fig.~\ref{Figure_01}(b)]
\cite{PhysRevLett.95.266802,PhysRevLett.99.256802,mele2010commensuration,bistritzer2011moire,Moon2013,koshino2015interlayer}.
At most wave vectors $\Vec{k}_0$ in the Brillouin zone,
these interactions are not very strong
because the involved monolayers states have different energies;
this is the reason why most parts of the band structure of twisted bilayers
are unchanged compared to the monolayer band structure (e.g., Fig.~3 in Ref.~\cite{Moon2013}).
At specific wave vectors, however,
the monolayer energy of one of the $|\tilde{\Vec{k}}_i\rangle$ becomes close to
that of $|\Vec{k}_0\rangle$ [solid line in Fig.~\ref{Figure_01}(b)],
and, hence, the resonant interaction at these points results in a band structure that is
very different from the monolayer band;
such an interaction forms either mini Dirac points or saddle point van Hove singularities.

At specific stacking configurations,
e.g., bilayers of hexagonal lattices stacked at a twist angle $\theta=30^\circ$ [Fig.~\ref{Figure_01}(a)]
or bilayers of square lattices at $\theta=45^\circ$,
the systems no longer has in-plane periodicity
but gains an $n$-fold quasicrystalline rotational symmetry
($n=12$ for the bilayers of hexagonal lattices and $n=8$ for the bilayers of square lattices)
\cite{stampfli,Ahn2018,suzuki2019ultrafast,takesaki2016highly,yao2018quasicrystalline,chen2016high,pezzini202030}.
Even in these van der Waals quasicrystals,
the states at most wave vectors show almost decoupled states
or a simple two-wave mixing, as shown in Fig.~\ref{Figure_01}(b).
At specific wave vectors, however,
a monolayer state in one layer
forms a resonant interaction with two states in another layer,
and each of these two states also forms
a resonant interaction with two states in the first layer.
And finally,
$n/2$  states in each layer
form a closed loop of resonant interactions
and exhibit the characteristic quasicrystalline band dispersion
\cite{moon2019quasicrystalline,PhysRevB.102.045113,PhysRevB.103.045408}.
The wave functions of these states have $n$-fold rotational symmetry,
which is incompatible with the periodicity.

However, the search for such a quasicrystalline resonant state
in twisted multilayers with more than two layers
is not straightforward.
For example, Fig.~\ref{Figure_01}(d) shows the
real-space lattice structures of three hexagonal lattices
stacked with $\theta = 20^\circ$ between the adjacent layers,
which shows a 18-fold rotational symmetry
if we disregard the difference between the vertical coordinates.
We can find a set of wave vectors,
of which Bloch states satisfy the Umklapp scattering (momentum conservation) condition
with respect to the interlayer interaction
and have the same monolayer state energy
[Fig.~\ref{Figure_01}(e), see Figs.~\ref{Figure_04}(f)-(j) for the actual wave vectors].
Thus, the set of such states
will form a closed loop of resonant interaction,
just like its bilayer counterpart.
Unlike the bilayer quasicrystals, however,
the part of the interaction path
that couples the states in distant layers (dashed lines)
is much weaker than the other parts of the path
that couple the states in adjacent layers (solid lines).
Thus, the resonant chain splits up into
a sequence of almost disconnected multi-atomic chains,
and it is unlikely that
such a state will satisfy the quasicrystalline symmetry.

On the other hand,
there has been a rapid progress in
the synthesis of Eshelby-twisted multilayers,
an infinite stack of atomic layers
with a constant twist angle $\theta$ between them \cite{liu2019helical}.
And theoretical investigation 
on such structures
revealed the overlap of the flat and dispersive bands
at small $\theta$ \cite{cea2019twists},
the $\theta$-dependent transitions
between type-I and type-II Weyl fermions \cite{PhysRevResearch.2.022010},
and chirality-specific nonlinear Hall effect \cite{PhysRevB.103.245206}.
These states are a natural generalization to three dimensions
of the non-resonant states in twisted hexagonal bilayers [Fig.~\ref{Figure_01}(b)] -
they have a $C_{3z}$ symmetry and corresponding angular quantum number $m=0,\pm1$
regardless of $\theta$,
due to the hexagonal symmetry of the moir\'{e} interference pattern.


It is, then, natural to ask
whether Eshelby-twisted multilayers can host
the resonant states
which arise from the interaction between the degenerate monolayer states
of the constituent layers.
In this paper,
we show that 
Eshelby-twisted multilayers composed of graphite,
a chiral twisted graphite (CTG),
with any $\theta$
can host such resonant states at specific points in the Brillouin zone
[Fig.~\ref{Figure_01}(f)].
Each resonant state exhibits
rich structures that differ from
the typical moir\'{e} band dispersion
at other wave vectors,
and its wave function has a screw rotational symmetry
which depends on $\theta$.
Each CTG has infinitely many distinct resonant states,
which can have up to four different screw symmetries.
For example,
a CTG with $\theta=30^\circ$,
of which lattice structure has a 12-fold screw symmetry,
has not only the resonant states with a 12-fold screw symmetry
but also the states with a 4-fold screw symmetry,
which is also incompatible with an in-plane periodicity.
Likewise, the CTGs with $\theta=12^\circ$ or $24^\circ$
of which lattice structure has a 30-fold screw symmetry,
has the resonant states with 5-, 10-, 15-, and 30-fold screw symmetries.
Interestingly, these resonant states have
a smaller ``reduced Brillouin zone'' than other non-resonant states in the same structure,
and the size of the reduced Brillouin zone
scales inversely with the order of the screw symmetry.
We also analytically show that the energies and wave functions of the resonant states
are written in a universal form of a one-dimensional chain 
regardless of $\theta$,
and reveal that the optical selection rules are described by
the difference between the quantum numbers of the wave functions associated with the screw symmetry.
Finally, we discuss
the uniqueness and existence of the ``exact center'' of the lattice 
and the self-similarity of the spatial distribution of the wave amplitudes.

\begin{figure}[ht]
\centering
\includegraphics[width=1.0\linewidth]{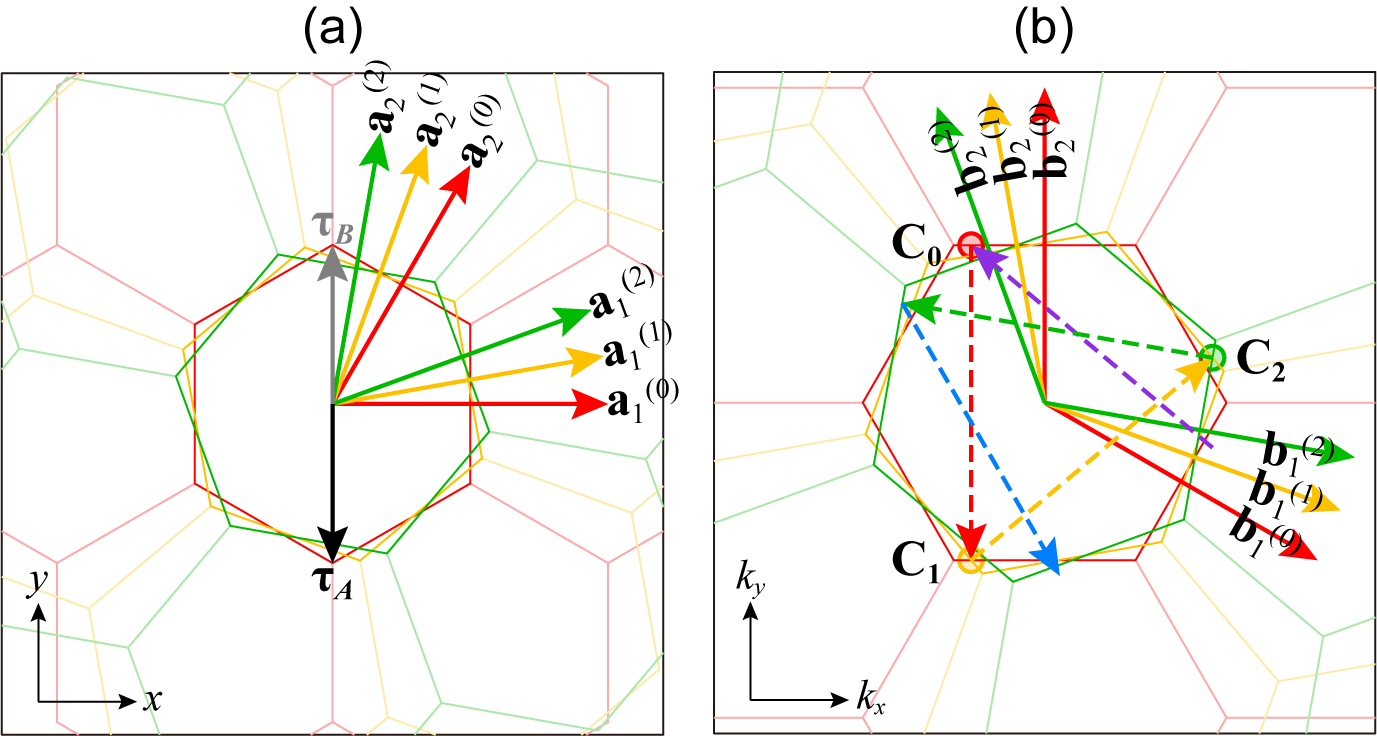}
\caption{
(a) Atomic structures of CTG with $\theta=10^\circ$,
showing only three constituent layers.
The red, yellow, green hexagons (arrows) show the unit cells [lattice vectors $\Vec{a}_i^{(l)}$ ($i=1,2$)] of the layers
with layer index $l=0,1,2$, respectively,
with $\GVec{\tau}_X$ ($X=A,B$) representing the sublattice coordinates of the $l=0$ layer.
(b) Extended Brillouin zone of the CTG in (a).
Each hexagon (arrow) shows the monolayer Brillouin zone [reciprocal lattice vectors
$\Vec{b}_i^{(l)}$
($i=1,2$)]
of each layer,
with $\Vec{C}_j$ and dashed lines representing the wave vector of the monolayer Bloch state in $j$-th layer
and the interlayer interactions
which forms the resonant interaction shown in Fig.~\ref{Figure_01}(f), respectively.
}
\label{Figure_02}
\end{figure}

The paper is organized as follows.
In Sec.~\ref{sec:atomic_structure_and_hamiltonian},
we present the atomic structure and
momentum-space tight-binding model
for CTGs.
We show the electronic structure
of CTGs calculated with full bases
and the emergence of unique band dispersion
near specific points in the Brillouin zone.
We discuss such ``resonant states'' in details in Sec.~\ref{sec:reson_states}.
We find the wave vectors of monolayer Bloch states
which form the resonant states (Sec.~\ref{sec:reson_cond}),
and show that there are infinitely many distinct resonant states,
which can have up to four different screw symmetries,
in each CTG (Sec.~\ref{sec:inf_many_reson_states}).
We build a Hamiltonian matrix of resonant states
with minimal
bases (Sec.~\ref{sec:H_ring}),
and analyze the band structure
and the size of the reduced Brillouin zone.
We discuss 
the optical selection rules of the resonant states 
and the exact center
in Secs.~\ref{sec:optical_selection_rule}
and \ref{sec:exact_center}, respectively.

\section{\label{sec:atomic_structure_and_hamiltonian}Atomic structure, Hamiltonian, and Electronic Structures of\\Chiral Twisted Graphite}

\subsection{\label{sec:atomic_structure}Atomic structure}


We consider an infinitely stacked CTG,
where each layer is rotated with respect to the lower layer by a fixed angle $\theta$.
Only the structures with $0<\theta\le30^\circ$ are distinct
due to the symmetry of the hexagonal lattice.
Such a structure is a helical lattice
which has the symmetry of an in-plane rotation by $\theta$
accompanied by a translation to the out-of-plane direction by the interlayer spacing,
i.e., a $\theta$ screw rotational symmetry,
and has no in-plane periodicity.
We set $xy$ coordinates parallel to the graphene layers
and $z$ axis perpendicular to the plane.
We define the atomic structure of the system
by starting from a nonrotated arrangement,
where the hexagon center of all the layers share the same
in-plane position $(x,y)=(0,0)$,
and the $A$-$B$ bonds are parallel to each other.
We choose $\Vec{a}_1 = a(1,0)$ and $\Vec{a}_2=a(1/2,\sqrt{3}/2)$
($a=0.246~\mathrm{nm}$)
as the primitive lattice vectors of graphene,
and $\GVec{\tau}_A=-\GVec{\tau}_1$ and $\GVec{\tau}_B=\GVec{\tau}_1$
[$\GVec{\tau}_1 = -(1/3)(\Vec{a}_1-2\Vec{a}_2)$]
as the coordinates of the $A$ and $B$ sublattices in the unit cell.
Then, we rotate the $l$-th layer by $l\theta$
and get the atomic positions of the $l$-th layer
\begin{equation}
\Vec{R}_{X}^{(l)}=n_1\Vec{a}_{1}^{(l)}+n_2\Vec{a}_{2}^{(l)}+\GVec{\tau}_X^{(l)}+ld\Vec{e}_z,
\end{equation}
where
$X=(A,B)$ denotes the sublattice index,
$n_i$ ($i=1,2$) are integers, 
$\Vec{a}_i^{(l)}=\mathcal{R}(l\theta) \Vec{a}_i$ and
$\GVec{\tau}_X^{(l)}=\mathcal{R}(l\theta) \GVec{\tau}_X$,
where $\mathcal{R}(l\theta)$ is a counterclockwise rotation by $l\theta$,
$d=0.335~\mathrm{nm}$ is the interlayer spacing between two adjacent layers and
$\Vec{e}_z$ is the unit vector normal to the layer.
The structure has a nonsymmorphic symmetry around the center of the rotation.
We show the layers with $l=0,1,2$ of CTG with $\theta=10^\circ$
in Fig.~\ref{Figure_02}(a).
We define the reciprocal lattice vectors $\Vec{b}_i^{(l)}$ of each layer
so as to satisfy
$\Vec{a}_{i'} \cdot \Vec{b}_i=2\pi\delta_{i'i}$ and
$\Vec{b}_i^{(l)}=\mathcal{R}(l\theta) \Vec{b}_i$,
and plot the reciprocal space of Fig.~\ref{Figure_02}(a)
in Fig.~\ref{Figure_02}(b).

Note that, although another similar material which constituent layers stacked with
an alternating twist angle,
$\theta$, $-\theta$, $\theta$, $-\theta$, $\dots$,
will also host an infinite chain of resonant interaction
at the middle of the two Dirac points of the neighboring layers
(i.e., $\bar{M}$ point of twisted bilayer graphene),
such structures are an intuitive expansion of
the periodic bilayer moir\'{e} superlattices.
Their electronic structures can be easily obtained by the conventional effective theory
which is based on the moir\'{e} periodicity.
Thus, we do not consider such 
structures with obvious in-plane periodicity in this work,
except the structure with $\theta=30^\circ$ which belongs also to the CTG class of materials.

No CTG has translational symmetry along the in-plane direction,
since the moire patterns defined by each pair of layers
are not commensurate.
In addition, most CTGs are not periodic along the $z$ axis as well
(hereafter ``incommensurate CTG'').
At specific $\theta$, however,
$l$-th layer can have an in-plane lattice configuration
the same as $(l+iN)$-th layers ($i\in\mathbb{Z}$)
for a fixed number $N\in\mathbb{Z}$.
Then the system is periodic with respect to the translation by $Nd\Vec{e}_z$
(hereafter ``commensurate CTG''),
and we can choose the $N$ successive layers as a primitive cell.
The allowed $\theta$ for commensurate CTG is
\begin{equation}
    \theta = 60^\circ \frac{M}{N}
\label{eq:theta}
\end{equation}
where $M \in \mathbb{Z}^+$, $M \le N/2$, and $\gcd(M,N)=1$.
$N$ and $M$ fully define the lattice geometry of commensurate CTG.
The structures with $N=2,3,4$,
where the primitive cell is bi-, tri-, quad-layer,
have only one lattice configuration
with $\theta=30^\circ, 20^\circ, 15^\circ$, respectively,
while that with $N=5$ has
two distinct configurations with $\theta=12^\circ$ and $24^\circ$,
which correspond to different $M$ (=1, 2).
The primitive cell with $N=2$ has the geometry
the same as that of the twisted bilayer graphene quasicrystal \cite{moon2019quasicrystalline,PhysRevB.103.045408}.

\begin{figure*}[ht]
\centering
\includegraphics[width=0.95\linewidth]{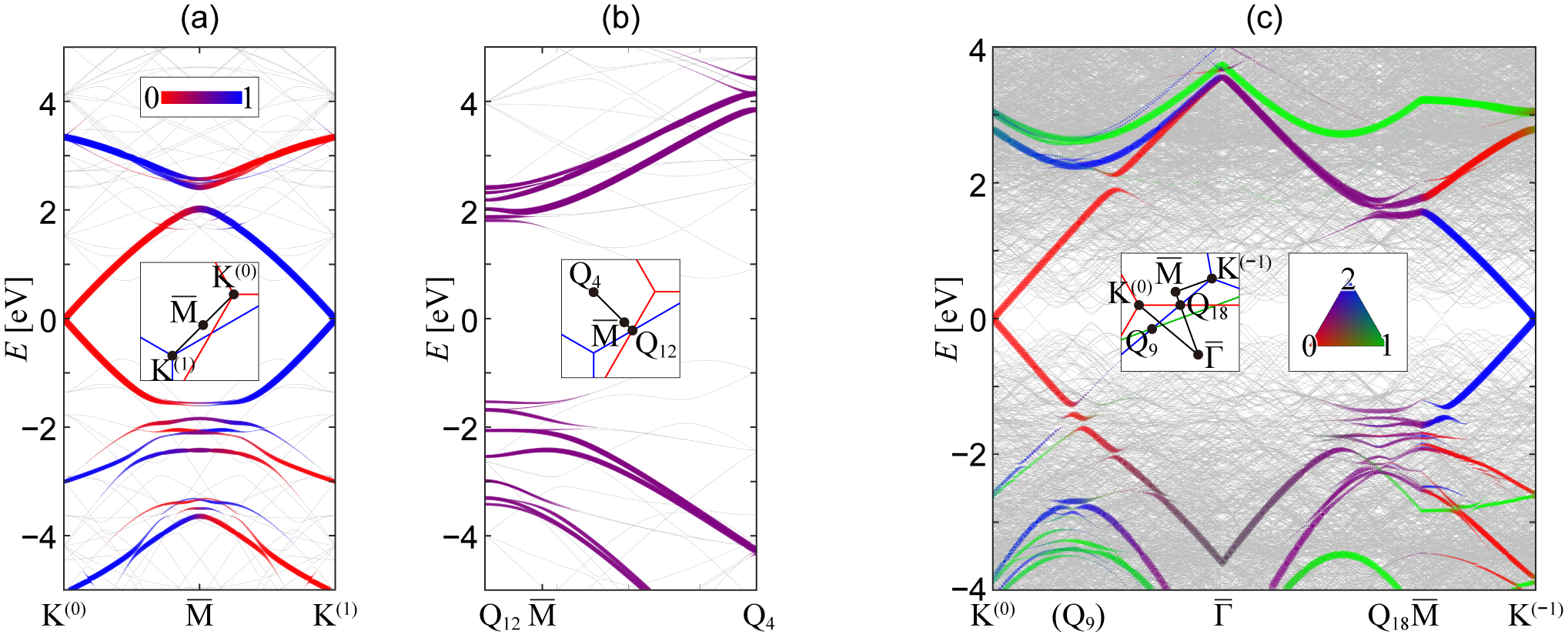}
\caption{
(a) and (b) Electronic band dispersion of a CTG with $N=2$ at $k_z=0$
plotted along (a) the line that connects the Dirac points
of the two adjacent layers,
and (b) the line that is perpendicular to (a).
The inset in the middle of each figure shows
the paths in the Brillouin zone used to plot the band dispersion.
$\Vec{Q}_{12}$ ($\Vec{Q}_4$) corresponds to one of the wave vectors
where the resonant states with a 12-fold (4-fold) screw rotational symmetry emerges
[see Figs.~\ref{Figure_04}(a) and (b)].
Gray lines show the bands calculated with the full bases model in Sec.~\ref{sec:dual_tb},
and the red and blue lines 
show the unfolded band dispersion
associated with the monolayer Bloch states in the layer $l=0,1$, respectively
[Eq.~\eqref{eq:spectral_function}].
The upper inset in (a) shows the mixing of the wave functions in
the layers with $l=0$ and 1.
(c) Plot similar to (a) and (b)
for a CTG with $N=3$
plotted along the paths shown in the left inset.
The line passes by (but not exactly through) $\Vec{Q}_{9}$ and passes through $\Vec{Q}_{18}$,
which are one of the wave vectors
where the resonant interaction forms states with a 9-fold and a 18-fold screw rotational symmetry
[see Figs.~\ref{Figure_04}(f) and (g)].
Red, green, blue lines show the band dispersion
associated with the monolayer Bloch states in the layer $l=0,1$, and 2 (or $-1$), respectively,
and the right inset shows the mixing between the wave functions
in the layers with $l$.
}
\label{Figure_03}
\end{figure*}

\subsection{\label{sec:dual_tb}Tight-binding model in momentum space}

We use a tight-binding model of carbon $p_z$ orbitals
to describe the electronic structure
of the general CTGs.
We define the Bloch state of the $l$-th layer
with the two-dimensional Bloch wave vectors $\Vec{k}$ as
\begin{equation}
	|\Vec{k},X^{(l)}\rangle = 
	\frac{1}{\sqrt{\tilde{N}}}\sum_{\Vec{R}_{X}^{(l)}} e^{i\Vec{k}\cdot\Vec{R}_{X}^{(l)}}
	|\Vec{R}_{X}^{(l)} \rangle
\end{equation}
in incommensurate CTGs and as
\begin{align}
	&|\Vec{k},X^{(l)}\rangle  \nonumber\\
	&= 
	\frac{1}{\sqrt{\tilde{N}}N_z}\sum_{i_z}\sum_{\Vec{R}_{X}^{(l+i_z N)}} e^{i\Vec{k}\cdot\Vec{R}_{X}^{(l+i_z N)}}
	e^{i k_z (l+i_z N) d}
	|\Vec{R}_{X}^{(l+i_z N)} \rangle
\end{align}
in commensurate CTGs.
Here $|\Vec{R}_{X}^{(l)} \rangle$ is the atomic orbital at the site
$\Vec{R}_{X}^{(l)}$,
$\tilde{N} = S_\mathrm{tot}/S$ is
the number of the graphene unit cells
with an area $S=(\sqrt{3}/2)a^2$
in the total system area $S_\mathrm{tot}$,
and $N_z$ and $k_z$ are the number of the primitive cells
and the Bloch wave number along the vertical direction, respectively.
The Brillouin zone along $\Vec{k}_z$ of 
commensurate CTG with $N$ layers in a primitive cell is
\begin{equation}
    -\frac{\pi}{Nd}\le k_z < \frac{\pi}{Nd},
    \label{eq:size_of_BZ_usual}
\end{equation}
owing to the periodicity along the vertical direction [Eq.~\eqref{eq:theta}];
this is analogous to the fact that
the electronic states in a rhombohedral graphite
are periodic with respect to $2\pi/(3d)$ shift of $k_z$ \cite{MCCLURE1969425}.
We use a two-center Slater-Koster parametrization \cite{slater_koster,moon2012energy}
for the transfer integral between any two $p_z$ orbitals,
\begin{equation}
-T(\Vec{R}) = 
V_{pp\pi}\left[1-\left(\frac{\Vec{R}\cdot\Vec{e}_z}{|\Vec{R}|}\right)^2\right]
+ V_{pp\sigma}\left(\frac{\Vec{R}\cdot\Vec{e}_z}{|\Vec{R}|}\right)^2,
\label{eq_slater_koster}
\end{equation}
where
$\Vec{R}$ is the relative vector between two atoms, and
\begin{eqnarray}
V_{pp\pi} &&=  V_{pp\pi}^0 e^{- (|\Vec{R}|-a/\sqrt{3})/r_0}, \nonumber\\
V_{pp\sigma} &&=  V_{pp\sigma}^0  e^{- (|\Vec{R}|-d)/r_0},    
\end{eqnarray}
$V_{pp\pi}^0 \approx -3.38~\mathrm{eV}$ \cite{comment_on_parameters},
$V_{pp\sigma}^0 \approx 0.48~\mathrm{eV}$,
and $r_0 \approx 0.0453\,\mathrm{nm}$ \cite{TramblydeLaissardiere2010,Moon2013}.
The total tight-binding Hamiltonian of CTG
is expressed as
\begin{equation}
    \mathcal{H} = \mathcal{H}_\mathrm{G} + \mathcal{U}
\end{equation}
where $\mathcal{H}_\mathrm{G}$ and $\mathcal{U}$ represent
the Hamiltonian for the intralayer and interlayer interaction, respectively.
The intralayer interaction in each layer is given by
\begin{align}
&\langle\Vec{k}',X'^{(l)} | \mathcal{H}_G  |\Vec{k},X^{(l)}\rangle =  h_{X'X}^{(l)}(\Vec{k})  \delta_{\Vec{k}', \Vec{k}}, \nonumber\\
&h_{X'X}^{(l)}(\Vec{k})  = \sum_{\Vec{L}^{(l)}}
-T(\Vec{L}^{(l)}+\GVec{\tau}_{X'X}^{(l)}) e^{-i\Vec{k}\cdot(\Vec{L}^{(l)}+\GVec{\tau}_{X'X}^{(l)})},
\end{align}
where $\Vec{L}^{(l)}$ is the lattice vectors of the $l$-th layer
and
$\GVec{\tau}_{X'X}^{(l)} = \GVec{\tau}_{X'}^{(l)}- \GVec{\tau}_{X}^{(l)}$.
And the interlayer matrix element between the layer $l$ and $l'$
is written as \cite{PhysRevLett.95.266802,PhysRevLett.99.256802,mele2010commensuration,bistritzer2011moire,Moon2013,koshino2015interlayer}
\begin{align}
& \langle \Vec{k}',X'^{(l')}| \mathcal{U}	|\Vec{k},X^{(l)}\rangle
=u_{X' X}^{(l',l)}
e^{i k_z (l-l') d}
\delta_{\Vec{k}+\Vec{G}^{(l)},\Vec{k}'+\Vec{G}^{(l')}}, \nonumber\\
& u_{X'X}^{(l',l)}= 
-\sum_{\Vec{G}^{(l)}}\sum_{\Vec{G}^{(l')}}
{t}(\Vec{k}+\Vec{G}^{(l)})
e^{-i\Vec{G}^{(l)}\cdot\mbox{\boldmath \scriptsize $\tau$}_{X}^{(l)}
	+i\Vec{G}^{(l')}\cdot\mbox{\boldmath \scriptsize $\tau$}_{X'}^{(l')}},
\label{eq_matrix_element_of_U}
\end{align}
where $\Vec{G}^{(l)}=m_1^{(l)} \Vec{b}_1^{(l)} + m_2^{(l)} \Vec{b}_2^{(l)}$ and
$\Vec{G}^{(l')}=m_1^{(l')} \Vec{b}_1^{(l')} + m_2^{(l')} \Vec{b}_2^{(l')}$
($m_1^{(l)}, m_2^{(l)}, m_1^{(l')}, m_2^{(l')} \in \mathbb{Z}$)
run over all the reciprocal points
of layer $l$ and $l'$, respectively.
Here
\begin{eqnarray}
{t}(\Vec{q}) = 
\frac{1}{S} \int
T(\Vec{r}+ z_{X'X}^{(l',l)}  \Vec{e}_z) 
e^{-i \Vec{q}\cdot \Vec{r}} d\Vec{r}
\label{eq_ft}
\end{eqnarray}
is the in-plane Fourier transform of the transfer integral,
where $z_{X'X}^{(l',l)} = (\GVec{\tau}_{X'}^{(l')}-\GVec{\tau}_{X}^{(l)})\cdot\Vec{e}_z$.
Since the interaction strength $T(\Vec{R})$ exponentially decays
with the interatomic distance,
the interlayer interaction is meaningful
only between the adjacent layers ($|l'-l|=1$).
Likewise, $|t(\Vec{q})|$ also exponentially decays
as $|\Vec{q}|$ increases.
Note that both $T(\textbf{R})$ and $t(\textbf{q})$
are isotropic along the in-plane direction,
i.e., $T(\Vec{R})=T(|\Vec{R}|)$ and $t(\Vec{q})=t(|\Vec{q}|)$,
if the two orbitals involved have the same magnetic quantum number, such as $p_z$ in this work
\cite{PhysRevB.103.045408}.

Since CTG
does not have
an in-plane periodicity that is common to the entire system,
one needs to find a general bases
which does not rely on such periodicity.
It is straightforward to show that
the Hamiltonian $\mathcal{H}$ spans the subspace
\begin{equation}
\{|\Vec{k},X^{(l)}\rangle ~| ~ \Vec{k}
=\Vec{k}_0+
\sum\limits_{l'\in \mathcal{L}^{\backslash(l)}} 
\sum\limits_{\Vec{G}^{(l')}} \Vec{G}^{(l')} \},
\label{eq:subspace}
\end{equation}
for any $\Vec{k}_0$ in the momentum space,
where $\mathcal{L}$ is $\mathbb{Z}$ for incommensurate CTG
and $\{0,1,...,N-1\}$ for commensurate CTG,
and $\mathcal{L}^{\backslash(l)}=\{0,1,\dots,N-1\} \setminus{\{l\}}$.
If we take each Bloch state as a ``site'',
the whole subspace can be recognized as a tight-binding lattice in the momentum space,
which is the dual counterpart of the original tight-binding
Hamiltonian in the real space \cite{moon2019quasicrystalline}.
In this momentum-space tight-binding model,
the hopping between different sites
(the interlayer interaction $\mathcal{U}$)
of van der Waals multilayers
is an order of magnitude smaller than the potential landscape (the band energies of the monolayers).
Thus, 
in a similar manner to the Aubry-Andr\'{e} model in one dimensional
real-space lattice under an incommensurate perturbation \cite{aubry1980analyticity},
the eigenfunctions in our model tend to be localized to a few sites in momentum space.
The analysis on 
the degree of the localization
in Ref.~\cite{moon2019quasicrystalline} shows
that most states in van der Waals bilayers are made up of
an interaction of 20 or fewer (in most cases, just two or three) monolayer states.
For most $\Vec{k}_0$ in CTG,
the length of such an interaction chain does not scale with the number of the constituent atomic layers owing to the mismatch between either the momentum or the monolayer state energies.
Thus, although the size of the subspace [Eq.~\eqref{eq:subspace}]
increases drastically with the size of $\mathcal{L}$,
we only need a limited
number of bases spanned from $|\Vec{k}_0,X^{(l)}\rangle$ of an arbitrary layer, chosen
by applying suitable cut-off to both $|t(\Vec{q})|$ and
the energy difference between the two monolayer states $\Delta E$,
to describe the electronic structures near $\Vec{k}_0$
for any practical calculation.
This is one of the largest merits of using the momentum-space model,
compared to the real-space model which requires infinitely many atomic orbital bases
to represent incommensurate systems.
We can, then, 
obtain the quasiband dispersion of the system
by plotting the energy levels against $\Vec{k}_0$.
Here the wave number $\Vec{k}_0$ works like the crystal momentum
for the periodic system, and so it can be called the quasi-momentum
for the current structures.



\subsection{Electronic structure}
\label{sec:electronic_structure}

Figures \ref{Figure_03}(a) and (b) show the band dispersion at $k_z=0$
in the extended Brillouin zone of the commensurate CTG with $N=2$,
calculated by the momentum-space tight-binding model;
(a) shows the dispersion
along the line $K^{(0)}-\bar{M}-K^{(1)}$,
where $K^{(l)}$ is the Dirac point of $l$-th layer
and $\bar{M}$ is the point in the middle of the two Dirac points,
while (b) shows the dispersion along
the line passes through $\bar{M}$ and perpendicular to $K^{(0)}-K^{(1)}$.
Since CTG does not have an in-plane periodicity,
it has an infinitesimal distinct Brillouin zone,
and the bands are replicated incommensurately into the extended Brillouin zone (thin gray lines).
We can reveal the distinct, unfolded band dispersion
at wave vector $\Vec{k}_0$ and energy $\varepsilon$
by the spectral function which is defined as
\begin{equation}
    A_l(\Vec{k}_0,\varepsilon)=\sum_{\alpha,X}  |\langle \alpha | \Vec{k}_0,X^{(l)} \rangle |^2 \delta(\varepsilon-\varepsilon_\alpha),
    \label{eq:spectral_function}
\end{equation}
where $|\alpha\rangle$ and $\varepsilon_\alpha$ are
the eigenstate and the eigenenergy, respectively.
The red and blue lines in Figs.~\ref{Figure_03}(a) and (b) show
$A_l$ for $l=0,1$, respectively.
The left panel shows that the Dirac cones at $K^{(0)}$ and $K^{(1)}$
interact with each other at $\bar{M}$
resulting in the complex structure which
is not observed
in the usual moir\'{e} superlattices with in-plane periodicity \cite{Moon2013}.
The right panel shows that
such a rich structure
stems from the multiple bands
at $\Vec{Q}_4$ and $\Vec{Q}_{12}$.
In Sec.~\ref{sec:bs_and_psi},
we will show that the band dispersion at $\Vec{Q}_4$ and $\Vec{Q}_{12}$
represents the resonant states with 4- and 12-fold screw rotational symmetry, respectively.

Figure \ref{Figure_03}(c) shows the band dispersion at $k_z=0$
of the commensurate CTG with $N=3$
along the line shown in the left inset;
this is the line $K^{(0)}-\bar{\Gamma}-\bar{M}-K^{(-1)}$
in terms of the high symmetry points of the moir\'{e} Brillouin zone between
the layers with $l=0$ and $-1$.
The red, green, blue lines show the $A_l$ for $l=0,1,-1$, respectively,
and the right inset shows their mixing.
Again, although the overall structure looks like the band dispersion
of usual moir\'{e} superlattices \cite{Moon2013},
due to the momentum mismatch at most wave vectors in quasicrystalline systems
\cite{Ahn2018,moon2019quasicrystalline},
we can see rich structures at $\Vec{Q}_9$ and $\Vec{Q}_{18}$,
which arise from the resonant states with
9- and 18-fold screw rotational symmetry, respectively
(see Sec.~\ref{sec:bs_and_psi}).

\section{\label{sec:reson_states}Resonant states in chiral twisted graphite}

\subsection{\label{sec:reson_cond}Resonant conditions}

\begin{figure*}[ht]
\centering
\includegraphics[width=0.95\linewidth]{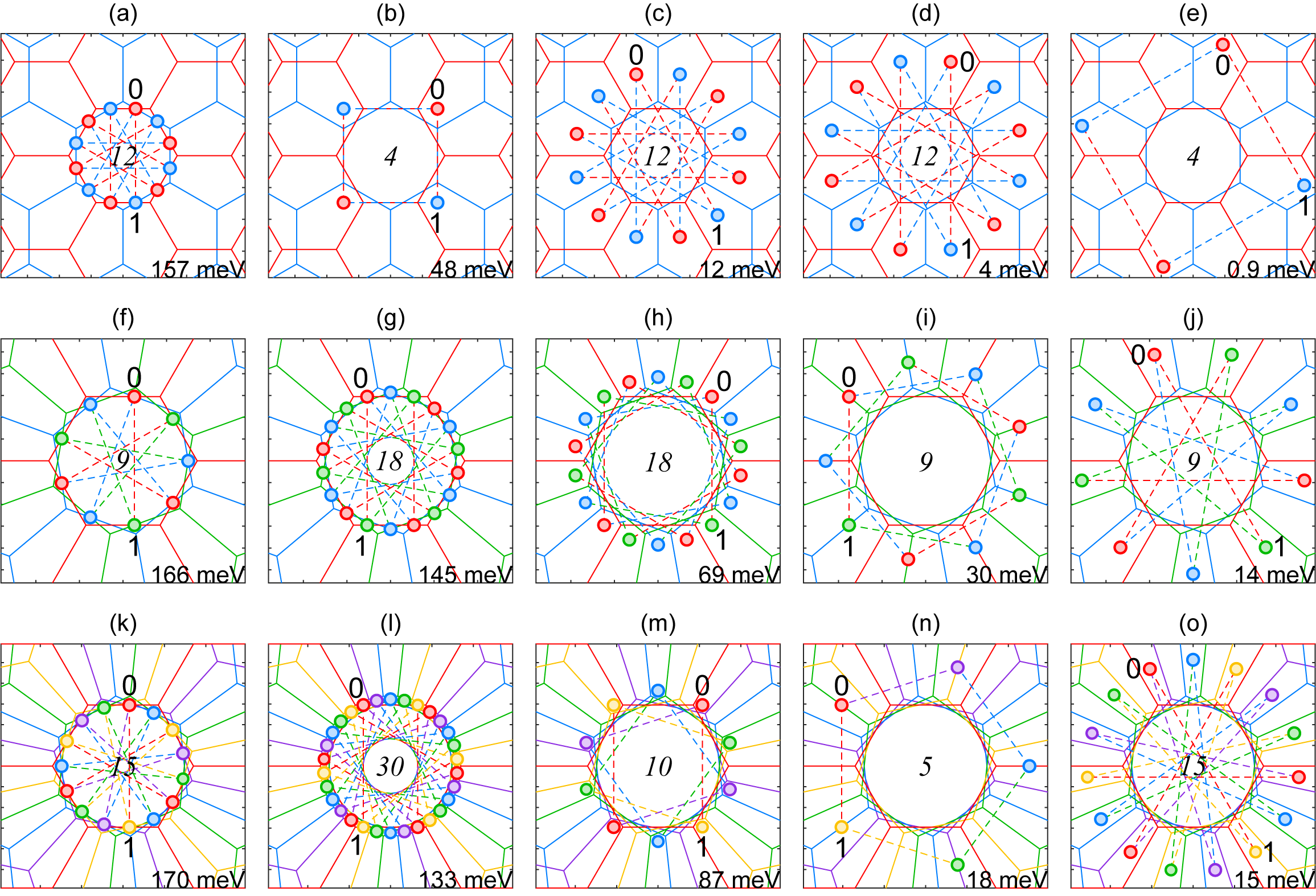}
\caption{
The sets of the wave vectors $\{\Vec{C}_j\}$ ($j=0,1,\dots,n-1$),
where the interaction between the monolayer Bloch states
form resonant states with $n$-fold screw rotational symmetry,
plotted in the extended Brillouin zone.
See Sec.~\ref{sec:reson_cond}.
(a)-(e) show the first five strongest resonant interactions
in a CTG with $N=2$
[$s$ is 3, 4, 3, 3, 4 for (a)-(e), respectively]
(f)-(j) in $N=3$
[$s$ is 3, 2, 4, 1, 3 for (f)-(j), respectively], and
(k)-(o) in $N=5$ and $M=1$
[$s$ is 3, 2, 4, 1, 3 for (k)-(o), respectively], 
respectively.
Each point and dashed line correspond to the monolayer Bloch state
and the change of momentum by interlayer interaction
associated with the layer of the same color.
Note that each dashed line corresponds to the reciprocal vector of each layer.
The italic number at the center shows the screw rotational symmetry $n$,
the energy at the bottom shows the interaction strength $t(\Vec{q})$,
``0'' and ``1'' represent $\Vec{C}_0$ and $\Vec{C}_1$, respectively,
and the vector $\Vec{C}_1-\Vec{C}_0$ shows the $\Vec{G}^{(0)}$ in each interaction.
Also note that CTGs with $N=5$ can have another configuration, $M=2$, which exhibits the resonant states distinct from those in (k)-(o) [e.g., the strongest resonant state has $n=30$ and $t(\Vec{q})=163~\mathrm{eV}$ (not shown)].
}
\label{Figure_04}
\end{figure*}

In Sec.~\ref{sec:electronic_structure},
we showed that the interlayer interaction at specific $\Vec{k}_0$ makes
a chain of
interaction
which strongly couples the degenerate monolayer states
of every constituent layers [Fig.~\ref{Figure_01}(f)].
The electronic structures of such states
are predominantly described by the interaction between these degenerate states.
To reveal such resonant interactions in incommensurate CTG, we need to find
a set of $\{\Vec{C}_j\}$ ($j \in \mathbb{Z}$), where
(i) all $|\Vec{C}_j,X^{(j)}\rangle$ are degenerate and
(ii) $|\Vec{C}_j,X^{(j)}\rangle$ in layer $j$
and $|\Vec{C}_{j+1},X^{(j+1)}\rangle$ in layer $j$+1
interact with each other by $\mathcal{U}$.
Then, due to the screw rotational symmetry of the system,
all $|\Vec{C}_j,X^{(j)}\rangle$ inevitably form an infinite chain of resonant interaction.
To satisfy (i) regardless of the band distortion, such as trigonal warping,
the relative position of $\Vec{C}_j$ to
the Dirac point of the $j$-th layer
($K^{(j)}$, either $K$ or $K'$)
must be the same in all the layers.
Since there are six Dirac points in each hexagonal Brillouin zone,
this condition requires $\Vec{C}_{j+1}-K^{(j+1)}=\mathcal{R}(\phi)(\Vec{C}_j-K^{(j)})$,
where $\mathcal{R}(\phi)$ is a counterclockwise rotation by
\begin{equation}
    \phi=\theta+60^\circ r \quad (r=0,1,\dots,5).    
    \label{eq:phi1}
\end{equation}
This can be further reduced to
\begin{equation}
    \Vec{C}_{j+1} = \mathcal{R}(\phi) \Vec{C}_j,
    \label{eq:(i)}
\end{equation}
since $K^{(j+1)}=\mathcal{R}(\phi) K^{(j)}$.
For (ii), $\Vec{C}_j$ and $\Vec{C}_{j+1}$ should satisfy
the generalized Umklapp scattering condition [Eq.~\eqref{eq_matrix_element_of_U}], i.e.,
\begin{equation}
    \Vec{C}_{j+1} + \tilde{\Vec{G}}^{(j+1)} = \Vec{C}_j + \Vec{G}^{(j)},
    \label{eq:(ii)}
\end{equation}
for the reciprocal lattice vectors of the $j$-th layer $\Vec{G}^{(j)}$ and
the ($j$+1)-th layer $\tilde{\Vec{G}}^{(j+1)}$, respectively.
Without loss of generality, however,
Eq.~\eqref{eq:(ii)} can be reduced to
\begin{equation}
    \Vec{C}_{j+1} = \Vec{C}_j + \Vec{G}^{(j)}
    \label{eq:(ii')}
\end{equation}
(see Appendix \ref{sec:App:umklapp}).
Then, $|\Vec{C}_j,X^{(j)}\rangle$ and $|\Vec{C}_{j+1},X^{(j+1)}\rangle$
interact with a magnitude of $|t(\Vec{q})|$,
where $\Vec{q} = \Vec{C}_j + \Vec{G}^{(j)}$
(= $\Vec{C}_{j+1}$).
We can obtain the set of $\{\Vec{C}_j\}$ which satisfies (i) and (ii)
by first finding $\Vec{C}_0$ that satisfies Eqs.~\eqref{eq:(i)} and \eqref{eq:(ii')},
\begin{align}
    \Vec{C}_0 &= [\mathcal{R}(\phi)-I_2]^{-1} \Vec{G}^{(0)} \nonumber\\
              &= \frac{-1}{2\sin(\phi/2)} \mathcal{R}(90^\circ-\frac{\phi}{2}) \Vec{G}^{(0)}
    \label{eq:C0}
\end{align}
where $I_2$ is a $2\times 2$ identity matrix,
for a reciprocal lattice vector of the 0-th layer $\Vec{G}^{(0)}$.
Then, due to the geometry,
$\Vec{G}^{(j)} = \mathcal{R}(j\phi) \Vec{G}^{(0)}$
for any $j$
is always a reciprocal lattice vector of the $j$-th layer
which satisfies Eq.~\eqref{eq:(ii')},
and all the states $|\Vec{C}_j,X^{(j)}\rangle$
with $\Vec{C}_j = \mathcal{R}(j\phi) \Vec{C}_0$
are degenerate.
Since all of these interactions have the same magnitude of coupling $t_0=t(|\Vec{q}|)$, where
\begin{equation}
    |\Vec{q}| = \frac{1}{2\sin (\phi/2)} |\Vec{G}^{(0)}|,
    \label{eq:|q|}
\end{equation}
as long as $t(\Vec{q})$ is isotropic (Sec.~\ref{sec:dual_tb}),
the set of the Bloch states $\{|\Vec{C}_j,X^{(j)}\rangle\}$
forms resonant interaction.

CTG becomes commensurate,
i.e., gains periodicity along the $z$ axis,
at specific $\theta$.
In addition to the resonant conditions (i) and (ii) for incommensurate CTG,
the $\{\Vec{C}_j\}$
of commensurate CTG
has to satisfy a periodic condition;
(iii) $|\Vec{C}_{j},X^{(j)}\rangle$ and $|\Vec{C}_{n+j},X^{(n+j)}\rangle$
with
some $n\in\mathbb{Z}$
are equivalent up to the $k_z$ phase.
This requires
that the lattice configuration of the ($n$+$j$)-th layer is the same as that of the $j$-th layer,
and also that $\Vec{C}_{n+j}=\Vec{C}_j$.
The former requires
\begin{equation}
    n = u N \quad (u\in\mathbb{Z}),
    \label{eq:n=uN}
\end{equation}
while the latter is equivalent to
\begin{equation}
    \sum_{j=0}^{n-1} \Vec{G}^{(j)}=\left[ \sum_{j=0}^{n-1} \mathcal{R}(j\phi) \right] \Vec{G}^{(0)} = 0,    
    \label{eq:(iii)}
\end{equation}
which requires
\begin{equation}
    \phi=360^\circ \frac{M'}{n},
    \label{eq:phi2}
\end{equation}
with $M' \in \mathbb{Z}$.
Without loss of generality,
we choose the smallest positive $n$ which satisfy (i)-(iii),
so that we describe
the shortest periodic unit of the interaction loop.
Then, we get
\begin{equation}
    Nr+M=\frac{6}{u}M'
    \label{eq:Nr+M}
\end{equation}
($n,u,M' \in \mathbb{Z}^+$)
from Eqs.~\eqref{eq:theta}, \eqref{eq:phi1}, \eqref{eq:n=uN}, and \eqref{eq:phi2},
and it is straightforward to show that $\gcd(u,M')=1$ and $\gcd(N,M')=1$
(see Appendix \ref{sec:App:N_u_Mp}).
Accordingly,
$\gcd(n,M')=1$ and
$u$ can only have a value in $\{1,2,3,6\}$.

As we will see later (Secs.~\ref{sec:H_ring} and \ref{sec:bs_and_psi}),
$\phi$ represents the angle of the screw rotational symmetry of each resonant state,
and $n$ for commensurate CTG represents the discrete screw rotational symmetry of the state.
In commensurate CTG, we can implement any $n$, except 1, 2, 3, 6,
with a suitable choice of the geometry ($N$, $M$) and $r$.

\subsection{\label{sec:inf_many_reson_states}Coexistence of infinitely many resonant states}

Above equations
indicate that there are infinitely many resonant states
in each CTG.
In a given geometry $\theta$,
different $r$ give
the resonant states with different $\phi$,
and in commensurate CTG,
the states with different $r$ may have different screw rotational symmetry $n$
[Eqs.~\eqref{eq:phi1} and \eqref{eq:Nr+M}].
We plot the five strongest resonant interactions
in CTGs with $N=2,3,5$ at the top, middle, and bottom panels in Fig.~\ref{Figure_04}.
We can see that a CTG with $(N,M)=(2,1)$ ($\theta=30^\circ$),
of which lattice configuration has a 12-fold screw rotational symmetry
(top panels),
can host not only the resonant states 
with a 12-fold rotational symmetry ($r=0,2,3,5$)
but also the states with a 4-fold rotational symmetry ($r=1,4$).
Besides, the system with $N=3$ ($N=4$) can host
the resonant states with $n=9,18$ ($n=8,24$),
while that with $N=5$ (bottom panels)
can have the states with $n=5,10,15,30$, and so on.
In addition,
even for a fixed $r$,
different $\Vec{G}^{(0)}$ make
the resonant states
with distinct $t(\Vec{q})$
which appear at different set of wave vectors $\{\Vec{C}_j\}$
in the Brillouin zone
[Eqs.~\eqref{eq:C0} and \eqref{eq:|q|}].
Despite the infinitely many resonant states,
however,
the number of distinct screw rotational symmetry $n$ in each commensurate CTG is finite;
this number 
is determined by the geometry $(N,M)$
and can be up to 4,
each of which corresponds to $u=1,2,3,6$,
at most.

\begin{figure*}[ht]
\centering
\includegraphics[width=0.9\linewidth]{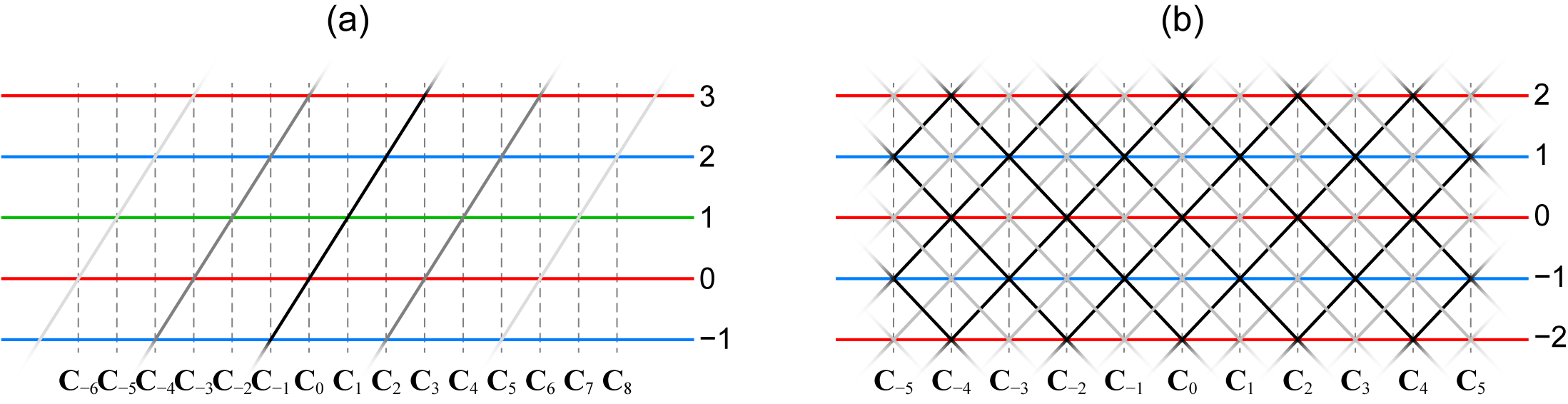}
\caption{
Schematic illustration of the resonant interaction chains
in CTGs with (a) $N=3$ and (b) $N=2$.
Each black or gray line shows the independent resonant interaction chain,
$\Vec{C}_j$ show the wave vectors of the Bloch states which form the resonant interaction,
and the numbers at the right show the layer indices.
}
\label{Figure_05}
\end{figure*}

Since $t(\Vec{q})$ decays exponentially as $|\Vec{q}|$ increases,
the interaction with the shorter $|\Vec{G}^{(0)}|$ exhibits
the stronger interaction between the monolayer Bloch states
[Eqs.~\eqref{eq:C0} and \eqref{eq:|q|}].
We, however, do not consider the interaction with $\Vec{G}^{(0)}=\Vec{0}$ in this work,
as such interaction merely represents the interaction between
the monolayer states at $\Gamma$ of every layers
which is common to multilayer systems with any $\theta$,
and the coupled states appear at the band edges of pristine graphene,
which are very far from the charge neutrality point.
Then, the strongest resonant interaction occurs
for the next shortest $|\Vec{G}^{(0)}|$,
i.e., $|\Vec{G}^{(0)}|=|\Vec{b}_i|$,
and $r=3$
(if $\theta=30^\circ$, both $r=2,3$ give the strongest interaction
). 
And without loss of generality,
we choose $\Vec{G}^{(0)}=-\Vec{b}_2^{(0)}$;
a different choice of $\Vec{G}^{(0)}$ having the same $|\Vec{G}^{(0)}|=|\Vec{b}_i|$
will only shift $\Vec{C}_{j_1}$ to $\Vec{C}_{j_2}$ in the same set of $\{\Vec{C}_j\}$,
and can be obtained by a similarity transformation.

\subsection{\label{sec:H_ring}Hamiltonian of resonant chain and size of reduced Brillouin zone}

The monolayer states $|\Vec{k}^{(j)},X^{(j)}\rangle$
($\Vec{k}^{(j)}=\Vec{C}_j+\Vec{k}_0$) in each layer $j$
form an infinitely long chain of resonant interaction.
Although the interlayer interaction [Eq.~\eqref{eq_matrix_element_of_U}]
couples these states also to other states,
the energies of such states
usually are very different from the energy of $|\Vec{k}^{(j)},X^{(j)}\rangle$,
except for accidental degeneracy.
Thus, we can get the details of the resonant states
by using these degenerated monolayer states as bases,
rather than using the full bases model described in Sec.~\ref{sec:dual_tb},
since the interaction to non-degenerate states
does not break the symmetry and degeneracy of the involved monolayer states
and merely shifts the resonant states by some, mostly small, constant energies
(see Appendix \ref{sec:outside}).
Note that
there are more resonant chains
which are associated with
$|\Vec{k}^{(cN)},X^{(0)}\rangle$ ($c\in\mathbb{Z}$)
[gray lines in Fig.~\ref{Figure_05}(a)],
in addition to the chain from
$|\Vec{k}^{(0)},X^{(0)}\rangle$
[black line in Fig.~\ref{Figure_05}(a)].
These chains are associated with
the $\Vec{G}^{(0)}$ different from the $\Vec{G}^{(0)}$ of
$|\Vec{k}^{(0)},X^{(0)}\rangle$
but having the same $|\Vec{G}^{(0)}|$.
Each interaction chain
makes the same band dispersion at different $\Vec{k}_0$.

To represent the Hamiltonian of the resonant states in this minimal bases model,
it is convenient to
use
new coordinate vectors
which are defined by the rotation by $\phi$, i.e.,
\begin{align}
    \Vec{a}_i^{(j)} &= \mathcal{R}(j\phi) \Vec{a}_i, \nonumber\\
    \GVec{\tau}_X^{(j)} &= \mathcal{R}(j\phi) \GVec{\tau}_X, \nonumber\\
    \Vec{b}_i^{(j)} &= \mathcal{R}(j\phi) \Vec{b}_i,
    \label{eq:newcoordvectors}
\end{align}
($i=1,2$, $X=A,B$),
instead of the vectors defined by the rotation by $\theta$ in Sec.~\ref{sec:atomic_structure}.
This is a unitary transformation which makes
the Hamiltonian a highly symmetric form
regardless of the distinct geometry and rotational symmetry of the resonant states
[$r$ in Eq.~\eqref{eq:phi1}].
Note that $|\Vec{k}^{(j)},X^{(j)}\rangle$
and $|\Vec{k}^{(j+cN)},X^{(j+cN)}\rangle$ ($c\in \mathbb{Z}$)
represent the Bloch states in the atomic layers with the same orientation,
but they are expressed by lattice vectors
rotated by $60 ^\circ crN$.
And the resonant states with different $r$
in the same lattice configuration
are expressed by coordinate vectors with different orientations.
In this choice of the coordinates,
all $\Vec{G}^{(j)}=m_1^{(j)}\Vec{b}_1^{(j)}+m_2^{(j)}\Vec{b}_2^{(j)}$
are represented by the same coefficients $m_1^{(j)}=m_1$, $m_2^{(j)}=m_2$.
Likewise, all $\Vec{C}_j=c_1^{(j)}\Vec{b}_1^{(j)}+c_2^{(j)}\Vec{b}_2^{(j)}$
are represented by the same coefficients
$c_1^{(j)}=c_1$, $c_2^{(j)}=c_2$,
where
\begin{equation}
    \begin{pmatrix}
      c_1\\c_2
    \end{pmatrix}
    = \left[ \tilde{\mathcal{R}}(\phi) - I \right]^{-1}
    \begin{pmatrix}
      m_1 \\ m_2
    \end{pmatrix}.
\end{equation}
Here,
\begin{equation}
    \tilde{\mathcal{R}}(\phi) = \frac{2}{\sqrt{3}}
    \begin{pmatrix}
      \cos(\phi-30^\circ) & -\sin \phi \\
      \sin \phi & \cos(\phi+30^\circ)
    \end{pmatrix}
\end{equation}
is a counterclockwise rotation by $\phi$
applied to the coefficients of
$\Vec{b}_1$ and $\Vec{b}_2$.

Then the Hamiltonian of any resonant interaction near $\Vec{k}_0=0$
of CTG
can be expressed by the Hamiltonian of 
one-dimensional chain,
\begin{align}
    &\hat{\mathcal{H}}_\mathrm{ring}(\Vec{k}_0) = \sum_{j \in \mathbb{Z}}
     \left\{ \psi_j^\dagger H^{(j)} \psi_j + [\psi_{j+1}^\dagger W \psi_j + \mathrm{H.c.}]
     \right\}, \nonumber\\
    &  H^{(j)}(\Vec{k}_0) = 
    \begin{pmatrix}
    h_{AA}^{(j)} & h_{AB}^{(j)}  \\
    h_{BA}^{(j)} & h_{BB}^{(j)}  \\
    \end{pmatrix}, \nonumber\\
    \quad
    &W =  \left \{
  \begin{aligned}
    &- t_0 
    \begin{pmatrix}
    \tilde{\omega}^* & \tilde{\omega}\\
    \tilde{\omega}^* & \tilde{\omega}
    \end{pmatrix} , && \text{incommensurate CTGs} \\
    &- t_0 
    \begin{pmatrix}
    \tilde{\omega}^* & \tilde{\omega}\\
    \tilde{\omega}^* & \tilde{\omega}
    \end{pmatrix} e^{-i k_z d}, && \text{commensurate CTGs}
  \end{aligned} \right.
    \label{eq_H_ring_incommensurate}
\end{align}
where $\psi_j^\dagger$ ($\psi_j$) is a creation (annihilation) operator
for $(|\Vec{k}^{(j)},A^{(j)}\rangle,|\Vec{k}^{(j)},B^{(j)}\rangle)$,
$h_{X'X}^{(j)}=h_{X'X}^{(0)}[\mathcal{R}(-j\phi) \Vec{k}_0 + \Vec{C}_0]$
in the new coordinate system,
$t_0=t(\Vec{C}_0)$,
where we neglect the $\Vec{k}_0$ dependence of the interlayer matrix element $t(\Vec{q})$,
and $\tilde{\omega}=\omega^{m_1-2m_2}$ ($\omega=e^{2\pi i/3}$).
Since $\hat{\mathcal{H}}_\mathrm{ring}$ is obviously
symmetric under the substitution of $j$ by $j+1$,
the resonant states are symmetric
under the screw rotation by $\phi$,
i.e., in-plane rotation by $\phi$
followed by a translation along $\Vec{e}_z$ by $d$,
regardless of the value of $k_z$.

In commensurate CTG, the resonant chain
has a periodic unit of $\{\Vec{C}_j\}$ ($j=0,1,\dots,n-1$).
Then the Hamiltonian of resonant states near $\Vec{k}_0=0$
can be expressed by an one-dimensional ring model,
\begin{align}
    &\mathcal{H}_{\rm ring}(\Vec{k}_0) \nonumber\\
    &= \begin{pmatrix}
    H^{(0)} & W^\dagger &&&& W \\
    W & H^{(1)} & W^\dagger \\
    & W & H^{(2)} & W^\dagger \\
    && \ddots & \ddots &\ddots \\
    &&& W & H^{(n-2)} & W^\dagger \\
    W^\dagger &&&& W & H^{(n-1)}
    \end{pmatrix},
    \label{eq_H_ring}
\end{align}
in the bases of
$(|\Vec{k}^{(j)},A^{(j)}\rangle, |\Vec{k}^{(j)},B^{(j)}\rangle)$,
$(j=0,1,\dots,n-1)$.
Again, since $\tilde{\mathcal{H}}_\mathrm{ring}$ is
symmetric under a rotation by a single span of the ring
(i.e., moving $\Vec{C}_j$ to $\Vec{C}_{j+1}$),
the resonant states is symmetric
with respect to the $n$-fold screw rotation
regardless of the value of $k_z$.

It should be emphasized that
the resonant states in commensurate CTG are
$u$ ($=n/N$)-fold degenerate along $k_z$
whereas the other, non-resonant states in the same system are not.
This becomes clear by a similarity transformation
of $\mathcal{H}_\mathrm{ring}$
\begin{align}
&\tilde{\mathcal{H}}_\mathrm{ring}(\Vec{k}_0) = U^{-1}\mathcal{H}_\mathrm{ring}(\Vec{k}_0)U \nonumber\\
&=
\begin{pmatrix}
H^{(0)} & \tilde{W}^\dagger &&&& \tilde{W} e^{-i n k_z d} \\
\tilde{W} & H^{(1)} & \tilde{W}^\dagger \\
& \tilde{W} & H^{(2)} & \tilde{W}^\dagger \\
&& \ddots & \ddots &\ddots \\
&&& \tilde{W} & H^{(n-2)} & \tilde{W}^\dagger \\
\tilde{W}^\dagger e^{i n k_z d} &&&& \tilde{W} & H^{(n-1)}
\end{pmatrix},
\label{eq_H_ring_tilde}
\\
&\tilde{W} = W e^{i k_z d} =  - t_0 
\begin{pmatrix}
\tilde{\omega}^* & \tilde{\omega}\\
\tilde{\omega}^* & \tilde{\omega}
\end{pmatrix},
\end{align}
with a transformation matrix
\begin{equation}
    U =\mathrm{diag}(1,e^{-i k_z d},e^{-2i k_z d},\dots,e^{-i (n-1) k_z d}) \otimes I_2.
\end{equation}
Equation \eqref{eq_H_ring_tilde} is periodic
with $k_z \rightarrow k_z + 2\pi/(nd)$,
so the resonant states with a $n$-fold rotational symmetry
are $u$-fold degenerate along the $k_z$ direction.
Such a repetition unit in the Brillouin zone
\begin{equation}
    -\frac{\pi}{nd} \le k_z < \frac{\pi}{nd}
    \label{size_of_BZ_QC}
\end{equation}
which is $u$ times smaller than
that of the general, non-resonant states in CTG
[Eq.~\eqref{eq:size_of_BZ_usual}]
can be regarded as a reduced Brillouin zone.
Moreover, the infinitely many resonant states in the same system
(Sec.~\ref{sec:inf_many_reson_states})
also have different reduced Brillouin zone size
if they have different $u$.

\begin{figure*}[ht]
\centering
\includegraphics[width=0.95\linewidth]{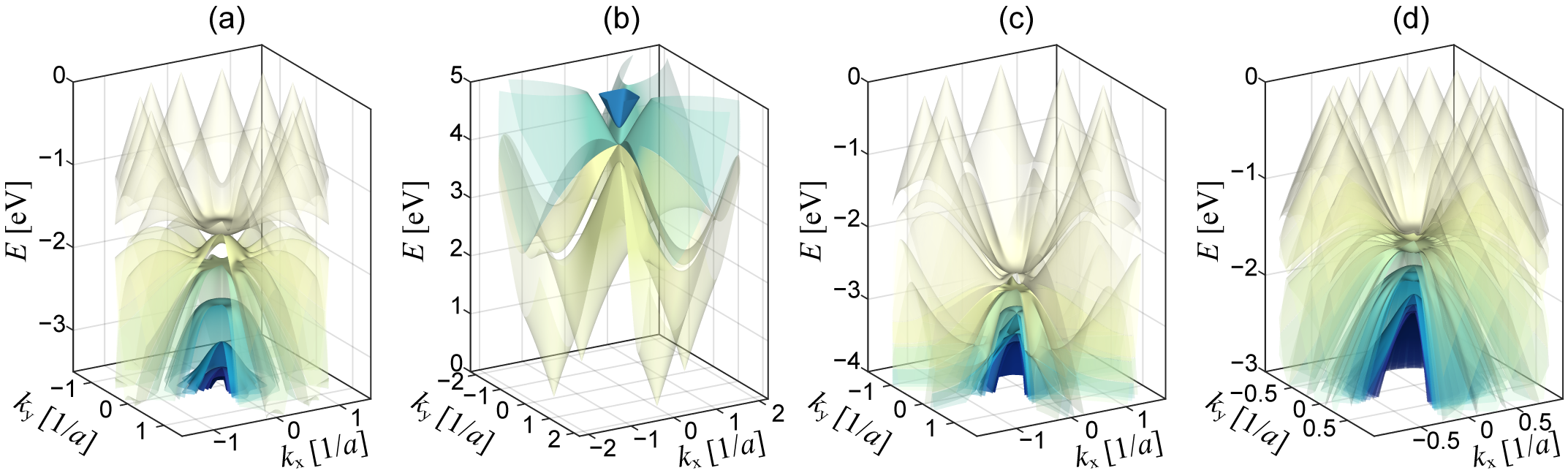}
\caption{
Electronic structure of the resonant states shown in Fig.~\ref{Figure_03}.
(a) and (b) show the valence bands of the states with $n=12$ 
and
the conduction bands with $n=4$
in a CTG with $N=2$, respectively.
(c) and (d) show the valence bands of the states with $n=9$ and 18
in a CTG with $N=3$, respectively.
Each band clearly shows the $n$-fold rotational symmetry.
}
\label{Figure_06}
\end{figure*}

The CTG with $N=2$ is, however, special in that
the monolayer state $|\Vec{C}_j,X^{(j)}\rangle$ in layer $j$
not only interacts with $|\Vec{C}_{j+1},X^{(j+1)}\rangle$ in layer $j+1$,
like the CTGs with other $N$ [Fig.~\ref{Figure_05}(a)],
but also interacts with $|\Vec{C}_{j+1},X^{(j-1)}\rangle$ in layer $j-1$
with the same momentum difference [Fig.~\ref{Figure_05}(b)].
Thus, $W$ should be replaced by
$W (1+e^{2 i k_z d})$ for $N=2$, and
such an extra phase changes the periodicity along $\Vec{k}_z$
of the resonant states 
from $2\pi/(nd)$ to $\mathrm{lcm}(2\pi/(nd),\pi/d)$ [Eq.~\eqref{eq_H_ring_tilde}].
Since only $n=4,12$ are allowed in $N=2$,
the size of the Brillouin zone along $\Vec{k}_z$
of the resonant states in $N=2$ is
the same as the Brillouin zone of the other, non-resonant states, $\pi/d$.
Except for the extra phase, $2\cos k_z d$, in the interlayer matrix elements,
$\mathcal{H}_\mathrm{ring}$ of $N=2$ is
the same as that of the twisted bilayer graphene quasicrystal in Ref.~\cite{moon2019quasicrystalline}.
Thus, the resonant states of the CTG with $N=2$
are natural generalization of
the quasicrystalline states of the twisted bilayer graphene quasicrystal
to the three-dimensional structures.
In CTG, however,
the wave functions of the upper and lower layers [Fig.~\ref{Figure_05}(b)]
interfere constructively at $k_z=c\pi/d$ ($c\in\mathbb{Z}$)
and double the magnitude of the interlayer interaction $W$ of $N=2$ 
compared to that of the twisted bilayer graphene quasicrystal.
At $k_z=\pi/(2d)+c\pi/d$, on the other hand,
the waves interfere destructively
and $W$ completely vanishes.
As a result, regardless of the matching of momentum and energies,
$|\Vec{C}_j,X^{(j)}\rangle$ at these $k_z$ are completely decoupled from each other,
and show the energies of the decoupled monolayer states.
Such a perfect constructive and destructive interference occur only at $N=2$.

\subsection{\label{sec:bs_and_psi}Band structures and wave functions}

We plot the band structure of the resonant states
with 12- and 4-fold screw rotational symmetry
in commensurate CTG with $N=2$
against $\Vec{k}_0$
in Figs.~\ref{Figure_06}(a) and (b), respectively,
and the resonant states with 9- and 18-fold screw rotational symmetries
in commensurate CTG with $N=3$
in Fig.~\ref{Figure_06}(c) and (d), respectively.
The dispersion in these plots
are consistent with the rich structure in Fig.~\ref{Figure_03}.
As shown in the previous section,
$N=2$ gives the dispersion the same as
that of the quasicrystalline twisted bilayer graphene,
but the interlayer interaction is doubled
at $k_z=c\pi/d$ ($c\in\mathbb{Z}$)
and vanishes in the middle between them.

At $\Vec{k}_0=\Vec{0}$, 
we can analytically obtain the energies of the resonant states (neglecting the constant energy)
\begin{widetext}
\begin{equation}
    E_m = -2t_0 \cos q_m \cos q_G \pm \sqrt{ (2t_0 \sin q_m \sin q_G)^2 + |h_0-2t_0 \tilde{\omega} \cos q_m|^2 }
    \label{eq:E(k0=0):incommensurate}
\end{equation}
for incommensurate CTGs,
\begin{equation}
    E_m(k_z) = -2t_0 \cos (q_z + q_m) \cos q_G \pm \sqrt{ (2t_0 \sin(q_z + q_m) \sin q_G)^2 + |h_0-2t_0 \tilde{\omega} \cos(q_z + q_m)|^2 }
    \label{eq:E(k0=0):commensurate}
\end{equation}
for commensurate CTGs with $N\ne 2$, and
\begin{equation}
    E_m(k_z) = -4t_0 \cos q_z  \cos q_m \cos q_G \pm \sqrt{ (4t_0 \cos q_z \sin q_m \sin q_G)^2 + |h_0-4t_0 \tilde{\omega} \cos q_z \cos q_m |^2 }
    \label{eq:E(k0=0):N=2}
\end{equation}
\end{widetext}
for commensurate CTGs with $N=2$,
as well as
the eigenvectors, i.e.,
the coefficients to the Bloch bases,
\begin{align}
    \Vec{v}_m^\pm &= (1/\sqrt{N'})(\dots,e^{i(j-1)q_m},e^{ijq_m},e^{i(j+1)q_m},\dots)^\mathrm{T} \nonumber\\
    &\otimes (c_{m,1}^\pm,c_{m,2}^\pm)^\mathrm{T}
    \label{eq:psi(k0=0)_incommensurate}
\end{align}
for incommensurate CTG and
\begin{align}
    \Vec{v}_m^\pm &= (1/\sqrt{n})(1,e^{iq_m},e^{2iq_m},\dots,e^{i(n-1)q_m})^\mathrm{T} \nonumber\\
    &\otimes (c_{m,1}^\pm,c_{m,2}^\pm)^\mathrm{T}
    \label{eq:psi(k0=0)}
\end{align}
for commensurate CTG,
by analytically diagonalizing $\tilde{\mathcal{H}}_\mathrm{ring}$.
Here $q_z=k_z d$,
$q_G = (2\pi/3) (m_1-2m_2)$,
$h_0 = h_{AB}^{(0)}(\Vec{C}_0)$,
$\pm$ corresponds to the conduction band and valence band, respectively,
$N'$ is the number of layers,
and see Appendix \ref{sec:c1_and_c2} for the expression of $c_{m,1}^\pm$ and $c_{m,2}^\pm$.
And $q_m=\phi m$
($m\in\mathbb{Z}$ for incommensurate CTG
and $m\in\{0,1,\dots,n-1\}$ for commensurate CTG)
is a Bloch number of the state which corresponds to the screw rotation by $\phi$.
Note that Eqs.~\eqref{eq:E(k0=0):incommensurate}-\eqref{eq:psi(k0=0)} are universal
to any CTG and any resonant states within,
although the parameters vary.
Each resonant interaction gives the eigenstates
twice as many as the number of distinct $q_m$,
i.e., infinite in incommensurate CTG and $2n$ in commensurate CTG.

\begin{figure}[ht]
\centering
\includegraphics[width=0.95\linewidth]{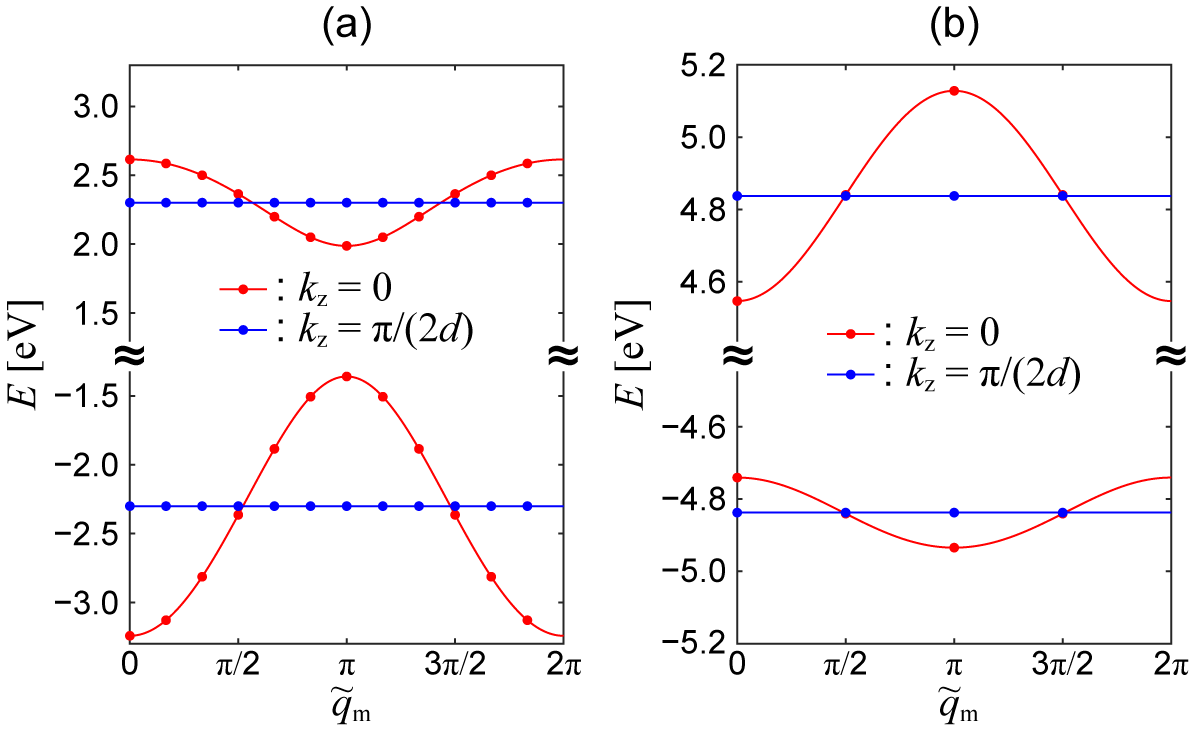}
\caption{
Energies of the resonant states $E_m$ in a CTG with $N=2$ [Eq.~\eqref{eq:E(k0=0):N=2}]
for $k_z=0$ (red) and $\pi/(2d)$ (blue)
plotted against a continuous (lines)
and discrete (dots) $\tilde{q}_m$ [Eq.~\eqref{eq:qm2}].
(a) and (b) show $E_m$ for the states with $n=12$ and 4, respectively.
(a) shows large dispersion in valence bands, while (b) shows larger dispersion in conduction bands,
which are consistent with Fig.~\ref{Figure_03}(b).
}
\label{Figure_07}
\end{figure}


Equation \eqref{eq:E(k0=0):N=2}
clearly shows that
the resonant states at $\Vec{Q}_{12}$
of the CTG with $N=2$
show a larger energy spacing in the valence band
than the conduction band,
just like the well-known
resonant states in
quasicrystalline twisted bilayer graphene,
while those at $\Vec{Q}_4$
show larger spacing in the conduction band
[Fig.~\ref{Figure_03}(b)].
We plot $E_m$ at $\Vec{Q}_{12}$ and $\Vec{Q}_4$
of a CTG with $N=2$ against
\begin{equation}
    \tilde{q}_m = \frac{2\pi}{n} m
    \label{eq:qm2}
\end{equation}
($m\in\{0,1,\dots,n-1\}$)
in Figs.~\ref{Figure_07}(a) and (b), respectively.
The red and blue lines show the dispersion
for $k_z=0$ and $k_z=\pi/(2d)$, respectively.
The set $\{\tilde{q}_m\}$ is congruent
to the set $\{q_m\}$, where
\begin{equation}
    q_m = \frac{2\pi M'}{n} m \ (=\phi m),
    \label{eq:qm1}
\end{equation}
modulo $2\pi$,
due to $\gcd(n,M')=1$.
In incommensurate CTGs, likewise,
the set $\{\tilde{q}_m\}$ [$\tilde{q}_m \equiv q_m$ (mod $2\pi$)]
is congruent
to $\{q_m\}$, 
since $\phi$ is incommensurate with $2\pi$.
As predicted in Sec.~\ref{sec:H_ring},
a CTG with $N=2$ experiences
a destructive interlayer interference
at $k_z=\pi/(2d)+c\pi/d$ ($c\in\mathbb{Z}$),
so $E_m$ at such $k_z$
are simply the monolayer energy at $\Vec{C}_0$
regardless of $\tilde{q}_m$.

\begin{figure*}[hbt]
\centering
\includegraphics[width=1.0\linewidth]{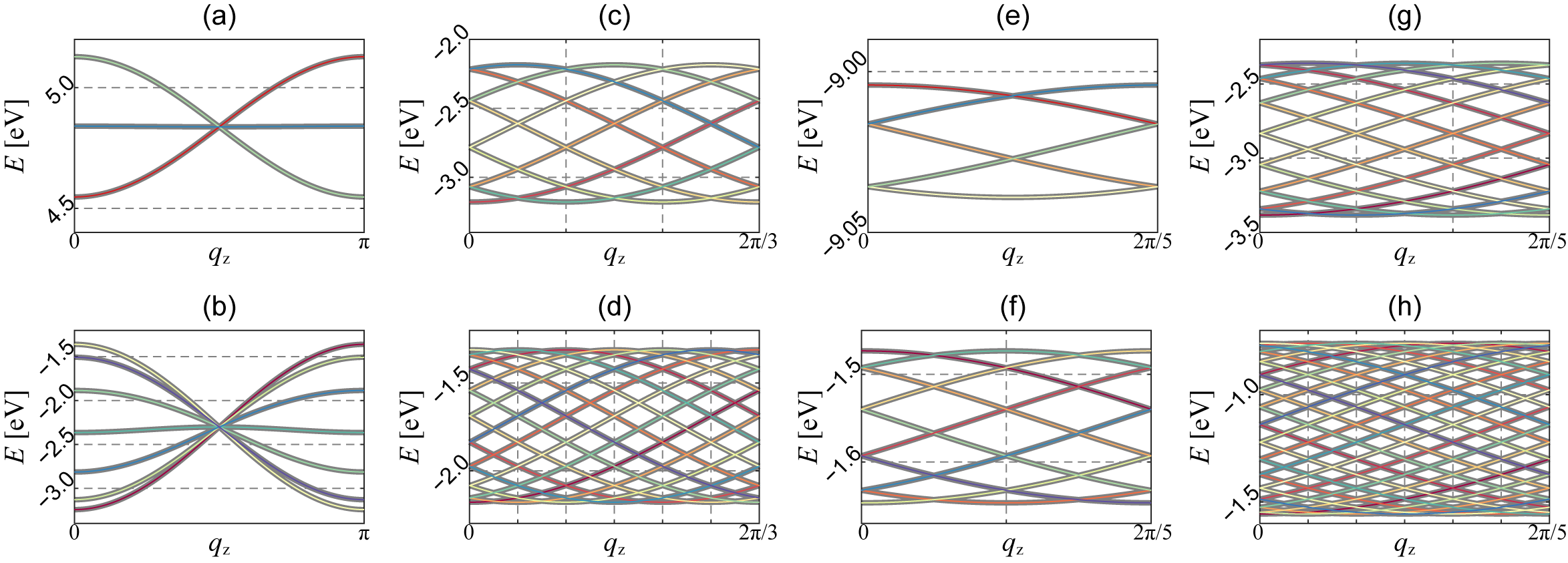}
\caption{
Energies of the resonant states $E_m$ plotted against $q_z(=k_z d)$.
(a) and (b) show the states with $n=4$ and 12 in $N=2$,
(c) and (d) show the states with $n=9$ and 18 in $N=3$,
(e), (f), (g), (h) show the states with $n=5, 10, 15, 30$ in $N=5$, respectively.
Lines with different colors show the states with different Bloch numbers $m$
associated with the screw rotation by $\phi$.
See text for the size
of the reduced Brillouin zone.
}
\label{Figure_08}
\end{figure*}

We plot $E_m$ of CTGs with $N=2,3,5$ against $q_z(=k_z d)$
in Fig.~\ref{Figure_08}.
Figures \ref{Figure_08}(a) and (b) show the states with $n=4$ and 12 in $N=2$,
(c) and (d) show the states with $n=9$ and 18 in $N=3$,
(e), (f), (g), (h) show the states with $n=5, 10, 15, 30$ in $N=5$, respectively.
Lines with different colors show the states with different Bloch numbers $m$
associated with the screw rotation by $\phi$.
It is clear that 
the reduced Brillouin zone of
the resonant states with a $n$-fold screw rotational symmetry along $k_z$ (vertical dashed lines)
is $n/N$ times smaller than 
the Brillouin zone $0\le k_z < 2\pi/(Nd)$ of the typical non-resonant states,
except for a CTG with $N=2$ (see Sec.~\ref{sec:H_ring}).

Equation \eqref{eq:qm2},
together with Eq.~\eqref{eq:E(k0=0):commensurate}
shows that the band energies of the resonant states
in commensurate CTGs with $N\ne 2$
exhibit a period of $2\pi/(nd)$ along $\Vec{k}_z$.
Thus, again,
the size of the reduced Brillouin zone of
resonant states
with a rotational quantum number of $n$
along $\Vec{k}_z$
[Eq.~\eqref{size_of_BZ_QC}]
is $u$ times smaller than that of the usual states [Eq.~\eqref{eq:size_of_BZ_usual}].
The proof in Sec.~\ref{sec:H_ring} is, however, more general,
since $\mathcal{H}_\mathrm{ring}(\Vec{k}_0)$ therein is valid even at $\Vec{k}_0\ne 0$.

\begin{figure*}[hbt]
\centering
\includegraphics[width=1.0\linewidth]{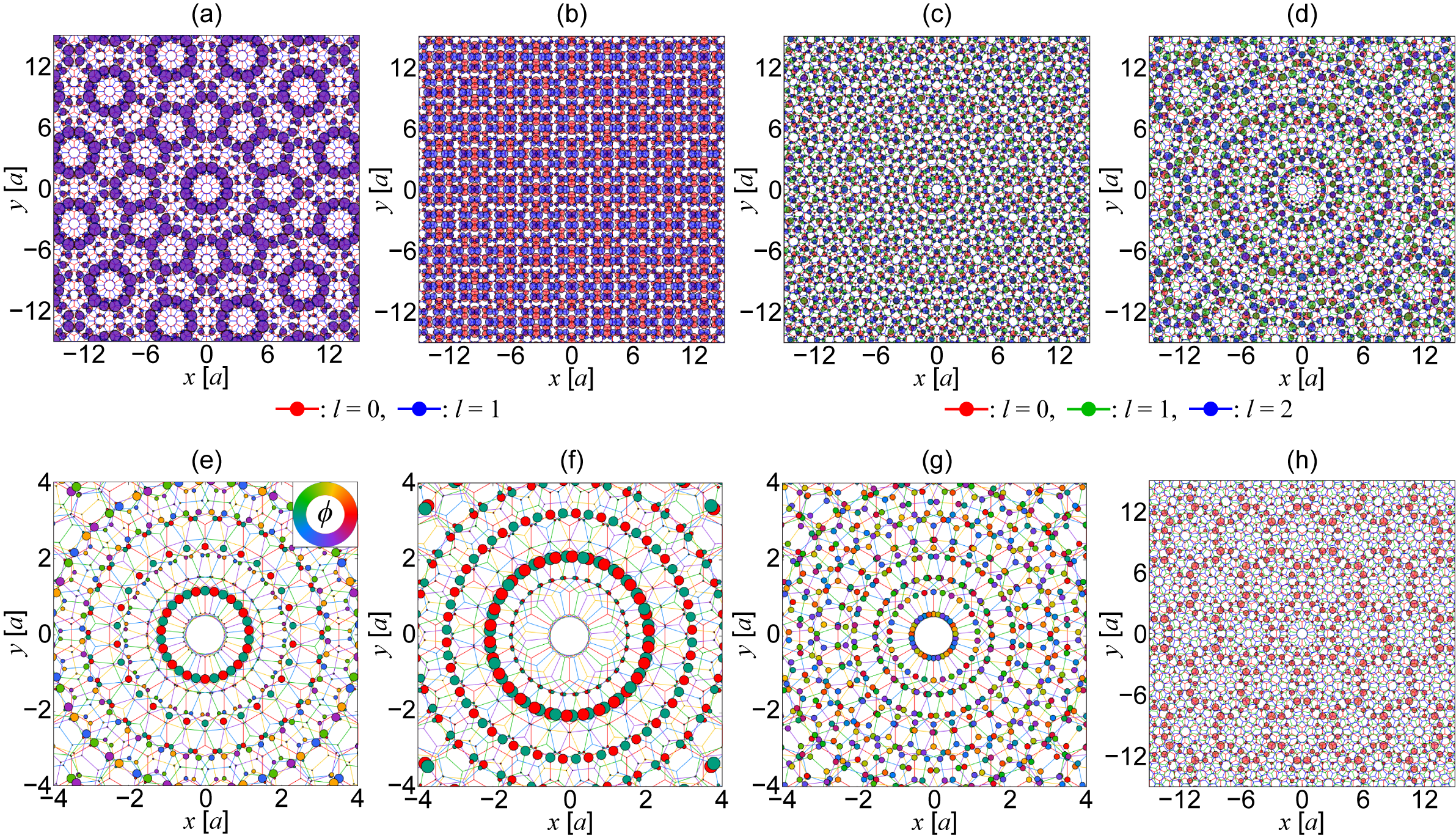}
\caption{
Spatial distribution of the wave functions of the resonant states
with (a) $n=12$, (b) $n=4$ in a CTG with $N=2$,
with (c) $n=9$, (d) $n=18$ in a CTG with $N=3$,
and with (e) $n=15$, (f) $n=30$, (g) $n=5$ in a CTG with $N=5$ and $M=1$,
each of which clearly show the $n$-fold rotational symmetry, respectively.
The area of the circle is proportional to the squared wave amplitude.
Each circle in (a)-(d) is colored according to the layer index $l$,
while that in (e)-(g) is colored by the phase of the wave function.
The inset in (e) shows the color wheel for the phase in (e)-(g).
(h) Plot similar to (d), but showing the wave functions
in a layer with $l=0$ only.
}
\label{Figure_09}
\end{figure*}

\begin{figure}[hbt]
\centering
\includegraphics[width=0.8\linewidth]{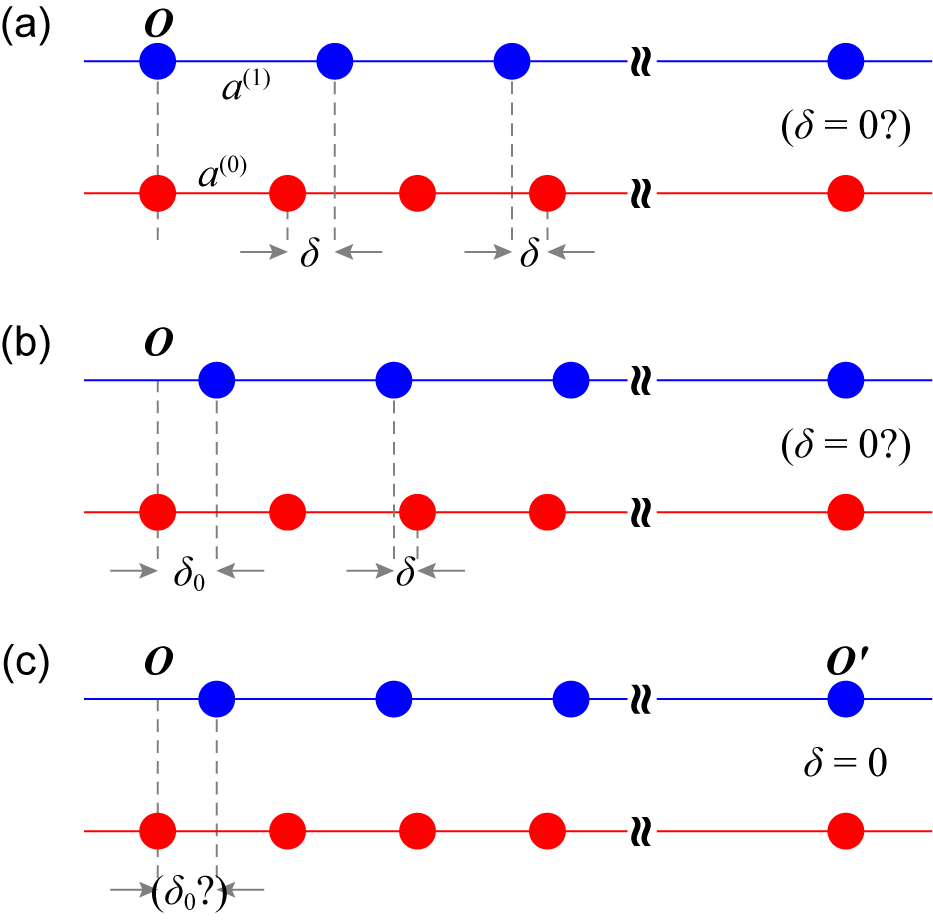}
\caption{
Lattice structures of a stack of two one-dimensional atomic layers
with the lattice constant of $a^{(l)}$ ($l=0,1$ is the layer index).
\Vec{O} represents the origin of the coordinate
and $\delta$ is the relative in-plane offset between the nearest atoms in the two layers at each lattice site.
(a) and (b) are the configuration where $\delta=0$ and $\delta_0$ at $\Vec{O}$, respectively,
and (c) is the configuration where $\delta=0$ at somewhere $\Vec{O}'$ in the system other than $\Vec{O}$.
}
\label{Figure_10}
\end{figure}

\begin{figure*}[hbt]
\centering
\includegraphics[width=0.8\linewidth]{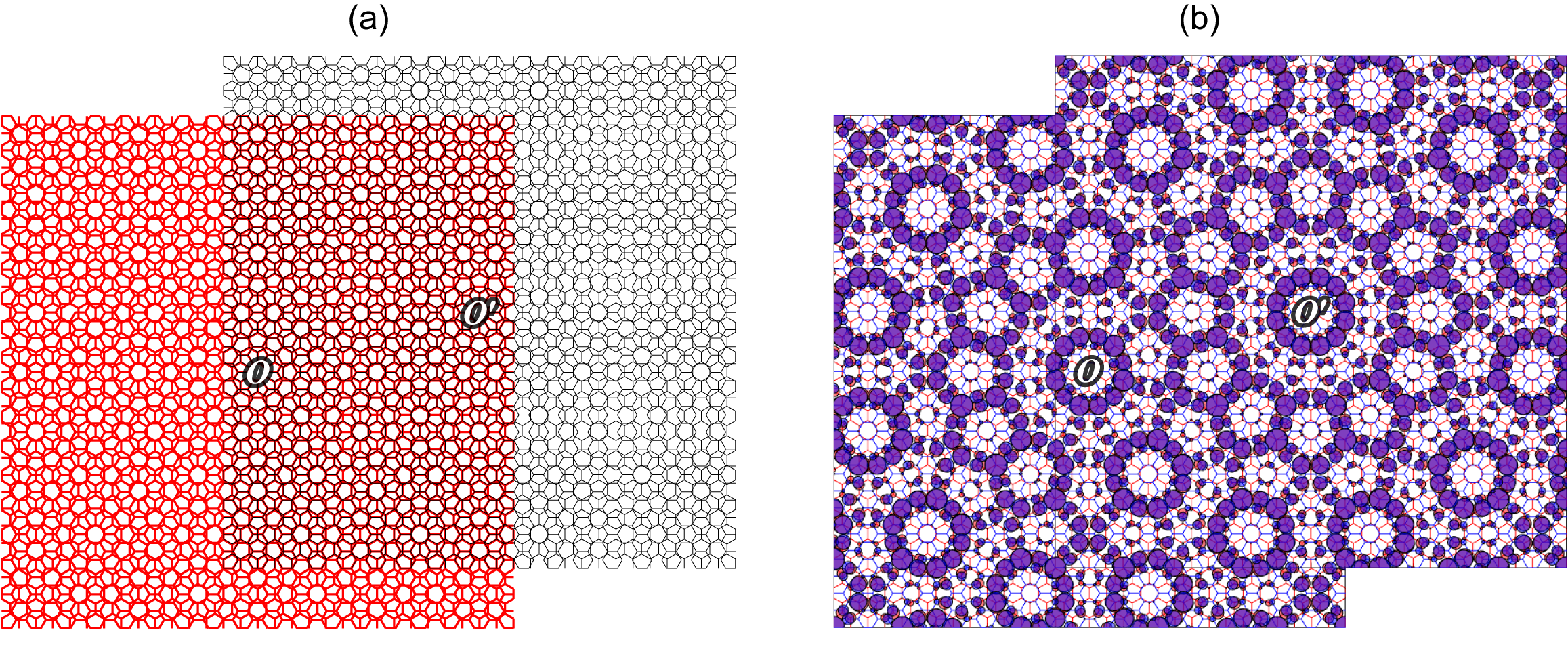}
\caption{
(a) Overlay of the original CTG lattice with $N=2$ (red lines)
and the same lattice translated by the vector between the exact center 
at the origin $\Vec{O}$
and similar centers $\Vec{O}'$ (gray lines).
The two lattices almost overlap in a finite-size region near the centers.
(b) Plot similar to (a) for the wave amplitudes of the resonant states with $n=12$ in the same system.
The two patterns almost seamlessly continue at the interface.
}
\label{Figure_11}
\end{figure*}

We plot the spatial distribution of the resonant states at $k_z=0$
in Fig.~\ref{Figure_09}.
Figures \ref{Figure_09}(a) and (b) [(c) and (d)] show the states
with a 12-fold and a 4-fold (a 9-fold and a 18-fold)
screw rotational symmetry in CTG with $N=2$ ($N=3$),
respectively.
And we plot the $l=0$ layer wave components 
of the wave functions in Fig.~\ref{Figure_09}(d)
to Fig.~\ref{Figure_09}(h).
Each circle in (a)-(d) is colored according to the layer index $l$,
and the area of the circle is proportional to the squared wave amplitude
at each atomic site.
We can clearly see that
the wave amplitude of the resonant states distribute selectively on a limited number of sites
satisfying the screw rotational symmetry.
Since $\mathcal{H}_\mathrm{ring}$ of the CTG with $N=2$
is identical to the Hamiltonian of the quasicrystalline states
in twisted bilayer graphene quasicrystal,
except for the extra phase from $k_z$,
a twisted bilayer graphene quasicrystal also exhibits
both the 12-fold and the 4-fold rotationally symmetric quasicrystalline states
at exactly the same $\{\Vec{C}_j\}$.
Note that neither 
of these resonant states
exhibit an in-plane periodicity,
since the linear combination of the wave vectors involved
is not commensurate with the periodicity
of underlying lattices.
In addition, we plot
the states with a 15-fold, a 30-fold, and a 5-fold symmetry in a CTG with $N=5$ and $M=1$
in Figs.~\ref{Figure_09}(e), (f), and (g), respectively,
where each circle is colored by the phase ($-\pi < \phi \le \pi$) of the wave functions
as shown in the color wheel in (e),
rather than the index of the layer.
Again, the spatial distribution of the phase also satisfies the rotational symmetry.

\subsection{\label{sec:optical_selection_rule}Optical selection rule}

In the coordinate system defined in Sec.~\ref{sec:H_ring},
the matrix elements of the velocity operators
$\hat{\Vec{v}}=(\hat{v}_x,\hat{v}_y)$
[$\hat{v}_\alpha=-(i/\hbar)[\alpha,\mathcal{H}]$, ($\alpha=x,y$)]
between the Bloch states of the $j$-th layer are written as
\begin{align}
&\langle\Vec{k}',X'^{(j)} | \hat{\Vec{v}} |\Vec{k},X^{(j)}\rangle = 
  \Vec{v}_{X'X}^{(j)}(\Vec{k})  \delta_{\Vec{k}', \Vec{k}}, \nonumber\\
&\Vec{v}_{X'X}^{(j)}(\Vec{k})  = \sum_{\Vec{L}^{(j)}}
  \Vec{R}_{X'X}^{(j)} T(\Vec{R}_{X'X}^{(j)}) 
  e^{-i\Vec{k}\cdot\Vec{R}_{X'X}^{(j)}} \nonumber\\
 & \qquad\quad\ \; = \mathcal{R}(j\phi) \sum_{\Vec{L}^{(0)}}
  \Vec{R}_{X'X}^{(0)} T(\Vec{R}_{X'X}^{(0)}) 
  e^{-i\Vec{k}\cdot\Vec{R}_{X'X}^{(0)}},
\end{align}
where $\Vec{R}_{X'X}^{(j)}=\Vec{L}^{(j)}+\GVec{\tau}_{X'X}^{(j)}=(R_{X'X,x}^{(0)}, R_{X'X,y}^{(0)})$.
Thus, we get
\begin{align}
    &v_{X'X}^{\pm,(j)}(\Vec{k}) \nonumber\\
    &= \langle \Vec{k},X'^{(j)} | \hat{v}_\pm | \Vec{k},X^{(j)} \rangle \nonumber\\
    &= e^{\pm ij\phi} \sum_{\Vec{L}^{(0)}}
  (R_{X'X,x}^{(0)}\pm i R_{X'X,y}^{(0)}) T(\Vec{R}_{X'X}^{(0)}) 
  e^{-i\Vec{k}\cdot\Vec{R}_{X'X}^{(0)}},
  \label{eq:v_pm_incommensurate}
\end{align}
for $\hat{v}_\pm=(\hat{v}_x\pm i \hat{v}_y)/\sqrt{2}$.

Equation \eqref{eq:v_pm_incommensurate},
together with Eqs.~\eqref{eq:psi(k0=0)_incommensurate} or \eqref{eq:psi(k0=0)},
reveals the optical selection rule
between the resonant states,
\begin{align}
    &\langle \Vec{v}_{m'}^{s'} | \hat{v}_\pm | \Vec{v}_m^s \rangle
    = V_{m'm}^{s's,\pm} \delta_{m',m\pm1}, \nonumber\\
    &V_{m'm}^{s's,\pm} =
        \begin{pmatrix}
            c_{m',1}^{s'} \\ c_{m',2}^{s'}
        \end{pmatrix}^\dagger
        \begin{pmatrix}
            v_{AA}^{\pm,(0)}(\Vec{C}_0) & v_{AB}^{\pm,(0)}(\Vec{C}_0) \\
            v_{BA}^{\pm,(0)}(\Vec{C}_0) & v_{BB}^{\pm,(0)}(\Vec{C}_0)
        \end{pmatrix}
        \begin{pmatrix}
            c_{m,1}^{s} \\ c_{m,2}^{s}
        \end{pmatrix},
\end{align}
where $s$ and $s'$ are $+$ for the conduction band and $-$ for the valence band.
Thus, the resonant states at $\Vec{k}_0=0$
exhibit a selection rule of $m\rightarrow m+1$ ($m\rightarrow m-1$)
for a right (left) circularly polarized light.




\subsection{\label{sec:exact_center}Presence of ``exact center'' and self-similarity of the lattice and wave amplitudes}

Each periodic lattice has a single periodic unit cell,
so, both their lattice structures and wave distributions can be identified
by spotting the single unit cell,
e.g., by using a scanning tunneling microscopy.
Given that both CTG with quasicrystalline twisted bilayer graphene do not have such periodicity,
it is natural to wonder how to identify the structure,
i.e., how to tell the geometry of the entire structure
from the local configuration.
For example,
a CTG may have the hexagonal centers of all the constituent layers
coincide (hereafter an ``exact center'')
at a specific 
position in the space;
in our choice of the geometry, the exact center appears at the origin.
If we can identify such an exact center,
we can easily identify the structure as a CTG with a specific $\theta$.
In experiments, however,
it is unlikely to find such a point in the entire system with an almost infinite size.

Before we proceed further,
it would be informative to investigate
(i) the uniqueness and (ii) the presence of a specific local atomic configuration,
such as the exact center,
in the entire system of incommensurate, non-periodic lattices.
Although here we consider a stack of one-dimensional atomic layers
to simplify the discussion,
it is straightforward to expand
these arguments to the stack of two-dimensional atomic layers
such as CTG and incommensurate twisted bilayer graphene with any $\theta$.

(i) \textit{Uniqueness}:
We first discuss the uniqueness of any local atomic configuration
in an incommensurate lattice.
We plot the atomic structure of one-dimensional incommensurate lattice
in Fig.~\ref{Figure_10}(a).
The red and blue lines show the layers with a layer index $l$ of 0 and 1, respectively,
of which lattice constant is $a^{(l)}$,
and the dots represent the atomic positions.
Let's define $\delta$ as the relative in-plane offset 
between the nearest atoms in the two layers
at each point.
In (a), we choose the configuration such that
the atoms of both layers coincide ($\delta=0$) at the origin $\Vec{O}$;
consider this as an ``exact center'' in the one-dimensional incommensurate lattices.
Then, the atomic coordinates of each layer are $\Vec{R}^{(l)}=n^{(l)} a^{(l)}$
($n^{(l)} \in \mathbb{Z}$).
The question about the uniqueness of the exact center asks
whether $\delta$ can be zero at any point other than $\Vec{O}$.
Since the lattice we consider is incommensurate,
we know that $a^{(1)}/a^{(0)} \not \in \mathbb{Q}$
and that $\delta=|n^{(1)}a^{(1)}-n^{(0)}a^{(0)}|$ cannot be zero at a point other than $\Vec{O}$.
Or, more simply,
if there are $n^{(0)}$ and $n^{(1)}$ ($n^{(0)}, n^{(1)} \ne 0$) 
which make $\delta=|n^{(1)}a^{(1)}-n^{(0)}a^{(0)}|$ to zero exist,
then any integer multiples of $(n^{(0)},n^{(1)})$ make $\delta=0$ as well,
and the system get a periodicity of $L=n^{(0)}a^{(0)}=n^{(1)}a^{(1)}$
which contradicts to the assumption that the system does not have a periodicity.
Thus, the exact center is, if it exists, unique in the entire space of an incommensurate lattice.
Note that this argument is valid not only for the exact center $\delta=0$,
but also for any $\delta=\delta_0$,
i.e., any local atomic configuration $\delta_0$ is unique in the entire system.

(ii) \textit{Presence}:
Now, we discuss the presence of any arbitrary local configuration
in incommensurate lattice.
We first investigate whether 
an incommensurate lattice with $a^{(0)}$ and $a^{(1)}$ the same as Fig.~\ref{Figure_10}(a)
but with an arbitrary $\delta_0\ne0$ at $\Vec{O}$
[Fig.~\ref{Figure_10}(b)]
can always host an exact center ($\delta=0$)
somewhere in the system.
This is equivalent to the question that
whether the $\delta$ at $\Vec{O}$
of a similar lattice which hosts an exact center somewhere in the system $\Vec{O}'$
[Fig.~\ref{Figure_10}(c)]
can have any arbitrary value $\delta_0$.
Measured from $\Vec{O}'$ (the exact center),
$\delta_0$ is given by $|n^{(1)}a^{(1)}-n^{(0)}a^{(0)}|$.
Note that $n^{(0)}, n^{(1)} \in \mathbb{Z}$,
while $\delta_0 \in \mathbb{R}$.
Both the sets $\{(n^{(0)}, n^{(1)})\}$ and $\{\delta_0\}$ are infinite sets,
but their cardinalities are different;
the cardinality of the set $\{(n^{(0)}, n^{(1)})\}$ is $\aleph_0$,
since there is a bijection $\mathbb{Z} \rightarrow \mathbb{Z}\times\mathbb{Z}$
and the cardinality of $\mathbb{Z}$ is $\aleph_0$,
while the cardinality of the set $\{\delta_0\}$ is $\aleph_1$.
Since $\aleph_0 < \aleph_1$,
we cannot host any arbitrary $\delta_0$ from $\Vec{O}'$ in Fig.~\ref{Figure_10}(c).
This means that
the presence of the exact center is not guaranteed
in incommensurate lattice with an arbitrary $\delta_0$ [Fig.~\ref{Figure_10}(b)].
Again, this argument is valid not only for the exact center $\delta=0$,
but also for any $\delta$,
i.e., not every local configuration appear in the entire system of an incommensurate lattice.

Above (i) and (ii) show that
it is unlikely to find the sites with high symmetry,
such as the exact center
which clearly show the characteristic lattice configuration,
by a scan of a finite range,
and such a site may not even exist.
Nevertheless, 
we can find sufficiently many similar sites,
since quasicrystalline lattices are self-similar at large scales
\cite{gardner1977extraordinary,levitov1988local,PhysRevX.6.011016}.
That is, any finite-size region appears infinitely many times on the space
in a non-periodic manner.
For example, Fig.~\ref{Figure_11}(a) shows that the exact center
of the CTG with $N=2$ appears multiples times in a space.
And the translation of the exact center
to one of those sites
maps the local finite-size regions before and after the translation.
Likewise, Fig.~\ref{Figure_11}(b) shows that
the spatial distribution of the wave amplitudes are also
coincide with such translation.
Note that, however, they are not exactly the same due to (i) above;
instead always have different surroundings.
In experiments, accordingly,
we will be able to see sufficiently many similar, but not exactly the same,
local configuration which shows the characteristic symmetry of the quasicrystalline configuration
in a finite-size region of the sample.



\section{Conclusions}

We investigated the electronic structures
of chiral, Eshelby-twisted van der Waals atomic layers,
in particular focusing on CTG.
We show that each CTG with any $\theta$
can host infinitely many resonant states
which arise from the interaction between the degenerate monolayer states
of the constituent layers.
Each resonant state has a screw rotational symmetry, which depends on $\theta$,
and each CTG has infinitely many distinct resonant states,
which can have up to four different screw symmetries.
The resonant states may have
a reduced Brillouin zone smaller than other non-resonant states in the same structure,
depending on the screw symmetry,
and exhibits
rich electronic structures
that differ from
the typical moir\'{e} band dispersion
at other wave vectors.

We derived the energies and wave functions of the resonant states
in a universal form of one-dimensional chain 
regardless of $\theta$.
The resonant states exhibit clear selection rules,
associated with the Bloch numbers of the screw rotational symmetry,
for circularly polarized light.
Finally,
we discuss the uniqueness and existence of the exact center of the CTG lattice
as well as the self-similarity of the wave amplitudes of the resonant states.
Although we explicitly used CTG in this work,
the methods introduced in this work
as well as the analysis on the research findings
can be easily expanded to
any Eshelby-twisted multilayers
as long as all the dominant interlayer interactions
occur between the atomic orbitals
that have the same magnetic quantum number.

\begin{acknowledgments}

P.M. acknowledges the support by
National Science Foundation of China (Grant No.~12074260),
Science and Technology Commission of
Shanghai Municipality (Shanghai Natural Science Grants,
Grant No.~19ZR1436400),
and the NYU-ECNU Institute of Physics at NYU Shanghai.
This research was carried out on the High Performance Computing
resources at NYU Shanghai.

\end{acknowledgments}

\appendix

\section{Umklapp scattering condition between $\Vec{C}_i$}
\label{sec:App:umklapp}

Suppose a set of $\{\Vec{C}'_j\}$ ($j \in \mathbb{Z}$) satisfies Eq.~\eqref{eq:(i)}, i.e.,
\begin{equation}
    \Vec{C}'_{j+1} = \mathcal{R}(\phi) \Vec{C}'_j.
    \label{eq:App:(i)}
\end{equation}
Then, the monolayer states of the $j$-th layer
with a Bloch wave vector $\Vec{k}=\Vec{C}'_j$
($|\Vec{C}'_j,X^{(j)}\rangle$)
and the state of the ($j$+1)-th layers
with a Bloch wave vector $\Vec{k}=\Vec{C}'_{j+1}$
($|\Vec{C}'_{j+1},X^{(j+1)}\rangle$)
interact with each other
when they satisfy the generalized Umklapp scattering condition
[Eq.~\eqref{eq_matrix_element_of_U}],
\begin{equation}
    \Vec{C}'_{j+1} + \tilde{\Vec{G}}'^{(j+1)} = \Vec{C}'_j + \Vec{G}'^{(j)},
    \label{eq:App:(ii)}
\end{equation}
for the reciprocal lattice vectors of the $j$-th layer $\Vec{G}'^{(j)}$ and
the ($j$+1)-th layer $\tilde{\Vec{G}}'^{(j+1)}$.
Then, the two monolayer states couple with
a magnitude of $|t(\Vec{q'})|$,
where $\Vec{q'}= \Vec{C}'_j + \Vec{G}'^{(j)} (= \Vec{C}'_{j+1} + \tilde{\Vec{G}}'^{(j+1)})$.

Due to the geometry, on the other hand,
$\mathcal{R}(-\phi) \tilde{\Vec{G}}'^{(j+1)}$ is always
a reciprocal lattice vector of the $j$-th layer.
Then, we can decompose $\Vec{G}'^{(j)}$
into a sum of another reciprocal lattice vectors, i.e.,
\begin{equation}
    \Vec{G}'^{(j)} = \Vec{G}^{(j)} + \Vec{G}''^{(j)},
    \label{eq:App:(ii)'}
\end{equation}
where
\begin{equation}
    \Vec{G}''^{(j)} = \mathcal{R}(-\phi) \tilde{\Vec{G}}'^{(j+1)}.
    \label{eq:App:(ii)''}
\end{equation}
With Eqs.~\eqref{eq:App:(i)}, \eqref{eq:App:(ii)'}, and \eqref{eq:App:(ii)''},
Eq.~\eqref{eq:App:(ii)} can be written as
\begin{equation}
    \mathcal{R}(\phi) \Vec{C}'_j + \mathcal{R}(\phi) \Vec{G}''^{(j)}
    = \Vec{C}'_j + \Vec{G}^{(j)} + \Vec{G}''^{(j)}.
\end{equation}
Then, we can define a new set of $\{\Vec{C}_j\}$,
by choosing $\Vec{C}_0 = \Vec{C}'_0 + \Vec{G}''^{(0)}$
and $\Vec{C}_j = \mathcal{R}(j\phi) \Vec{C}_0$,
which satisfies Eq.~\eqref{eq:(i)}
and a simplified Umklapp scattering condition,
\begin{equation}
    \Vec{C}_{j+1} = \Vec{C}_j + \Vec{G}^{(j)},
    \label{eq:App:(ii)'''}
\end{equation}
without loss of generality.
Since $\Vec{C}_j$ and $\Vec{C}'_{j}$ 
($\Vec{C}_{j+1}$ and $\Vec{C}'_{j+1}$) are equivalent
in the Brillouin zone of the $j$-th [($j$+1)-th] layer,
and the magnitude of the interaction in Eq.~\eqref{eq:App:(ii)'''}
is $|t(\Vec{q})|$, where
$\Vec{q} = \Vec{C}_j + \Vec{G}^{(j)} = \Vec{q}'$ from Eq.~\eqref{eq:App:(ii)'},
Eq.~\eqref{eq:App:(ii)'''} describes the interaction
exactly the same as Eq.~\eqref{eq:App:(ii)}
but in a much simpler form with $\tilde{\Vec{G}}'^{(j+1)}=0$.

\section{Parameters describing geometry and resonant states}
\label{sec:App:N_u_Mp}

We choose the smallest possible $n=u N$ ($u\in\mathbb{Z}^+$)
which satisfies (i)-(iii) in Sec.~\ref{sec:reson_cond},
so that we describe the shortest periodic unit of the interaction loop.
The condition (iii), i.e.,
$|\Vec{C}_{n+j},X^{(n+j)}\rangle = |\Vec{C}_{j},X^{(j)}\rangle$
up to the $k_z$ phase,
requires
\begin{equation}
    \phi=360^\circ \frac{M'}{n}=360^\circ \frac{M'}{u N},
    \label{eq:app:phi2}
\end{equation}
with $M' \in \mathbb{Z}^+$.
If $\gcd(u,M')=c$ with $c\ne 1$,
then there exist $\bar{M}'$ and $\bar{u}$ such that
$M'=c \bar{M}'$ and $u=c \bar{u}$, and
Eq.~\eqref{eq:app:phi2} becomes
$\phi=360^\circ \bar{M}'/ (\bar{u}N)$.
This indicates that
there exists an interaction chain with a length of $\bar{n}=n/c$
which satisfies (iii) before the $n$ scattering.
This contradicts to the initial assumption that
$n$ is the smallest integer satisfying (i)-(iii), so we get
\begin{equation}
    \gcd(u,M')=1.
    \label{eq:app:u_M'}
\end{equation}
Then, Eq.~\eqref{eq:Nr+M} 
\begin{equation}
    Nr+M=\frac{6}{u}M'
    \label{eq:app:Nr+M}
\end{equation}
and Eq.~\eqref{eq:app:u_M'}
show that $u$ can only have a value in $\{1,2,3,6\}$.

Suppose that there is $c\ne 1$ for $\gcd(Nr+M,N)=c$.
Then, there exist $\bar{N}$ and $\bar{M}$ such that
$N=c\bar{N}$ and $M=c\bar{M}$,
which contradicts to Eq.~\eqref{eq:theta}.
Thus, $\gcd(Nr+M,N)=1$,
and from Eq.~\eqref{eq:app:Nr+M} and $u=\{1,2,3,6\}$,
we get $\gcd(N,M')=1$.
And accordingly,
we also get $\gcd(n,M')=1$.


\section{Interaction to the states outside the resonant chain}
\label{sec:outside}
Each monolayer Bloch state in CTG
interacts with infinitely many other states in other layers
when they satisfy the generalized Umklapp scattering condition
[Eq.~\eqref{eq_matrix_element_of_U}].
The energies of such states usually have
various monolayer state energies,
except for accidental degeneracy.
In Sec.~\ref{sec:reson_states} and Fig.~\ref{Figure_04}, however,
we showed that the monolayer states at specific wave vectors
$\{\Vec{C}_j\}$
are degenerate and form the chains of resonant interaction,
which predominantly describe the electronic structures
near these wave points.
Then, it is natural to ask whether the interaction with the states
outside the resonant chain
breaks the resonant states
by lifting the degeneracy or symmetry
of the states forming the chain.

In Sec.~\ref{sec:offresonance:general},
we will first show that such additional interactions
in general CTGs with any $N$
do not lift the degeneracy of the constituent states,
since the interaction environment, i.e., surrounding,
of $\Vec{C}_{j}$ in the $j$-th layer is identical to the surrounding
of $\Vec{C}_{\tilde{j}}$ in the $\tilde{j}$-th layer ($\tilde{j} \ne j$).
Then, in Sec.~\ref{sec:offresonance:N=2}, we will 
visualize the resonant interactions and their surroundings
in a CTG with $N=2$
by using the momentum-space tight-binding model,
and also show the coupling between distinct resonant chains.

\subsection{General CTG}
\label{sec:offresonance:general}
Suppose a set of $\{\Vec{C}_j\}$ which makes up the resonant chain.
Then, each monolayer Bloch state $|\Vec{C}_j,X^{(j)}\rangle$
in $j$-th layer
not only interacts with the other $|\Vec{C}_{j'},X^{(j')}\rangle$
($j'\ne j$),
but also interacts with every monolayer states
$|\Vec{k}^{(j')},X'^{(j')}\rangle$
when they satisfy the Umklapp scattering condition
[Eq.~\eqref{eq_matrix_element_of_U}],
\begin{equation}
    \Vec{C}_j+\Vec{G}^{(j)}=\Vec{k}^{(j')}+\Vec{G}^{(j')},
    \label{eq:si:umklapp}
\end{equation}
for the reciprocal lattice vectors of the $j$-th layer
$\Vec{G}^{(j)}$ and the $j'$-th layer $\Vec{G}^{(j')}$.
The interlayer matrix element between those states
is written as
\begin{align}
    &\langle \Vec{k}^{(j')},X'^{(j')} | \mathcal{U} |
    \Vec{C}_j,X^{(j)} \rangle \nonumber\\
    &=-t(\Vec{q}) \, e^{i k_z (j-j') d} \,
    e^{-i\Vec{G}^{(j)}\cdot\mbox{\boldmath \scriptsize $\tau$}_{X}^{(j)}
	+i\Vec{G}^{(j')}\cdot\mbox{\boldmath \scriptsize $\tau$}_{X'}^{(j')}},
	\label{eq:si:matrix_element_of_U}
\end{align}
where
\begin{equation}
    \Vec{q}=\Vec{C}_j+\Vec{G}^{(j)}.
    \label{eq:si:q}
\end{equation}
Likewise, for $|\Vec{C}_{\tilde{j}},X^{(\tilde{j})}\rangle$
with $\tilde{j}\ne j$,
Eqs.~\eqref{eq:si:umklapp}-\eqref{eq:si:q}
become
\begin{align}
    &\Vec{C}_{\tilde{j}}+\Vec{G}^{(\tilde{j})}
    =\Vec{k}^{(\tilde{j}')} + \Vec{G}^{(\tilde{j}')}, 
    \label{eq:si:umklapp'}
    \\
    &\langle \Vec{k}^{(\tilde{j}')},X'^{(\tilde{j}')} | \mathcal{U} |
    \Vec{C}_{\tilde{j}},X^{(\tilde{j})} \rangle \nonumber\\
    &=-t(\Vec{\tilde{q}}) \, e^{i k_z (\tilde{j}-\tilde{j}') d} \,
    e^{-i\Vec{G}^{(\tilde{j})}\cdot\mbox{\boldmath \scriptsize $\tau$}_{X}^{(\tilde{j})}
	+i\Vec{G}^{(\tilde{j}')}\cdot\mbox{\boldmath \scriptsize $\tau$}_{X'}^{(\tilde{j}')}}, 
	\label{eq:si:matrix_element_of_U'}
	\\
	&\Vec{\tilde{q}}=\Vec{C}_{\tilde{j}}+\Vec{G}^{(\tilde{j})}
	\label{eq:si:q'}.
\end{align}
By comparing the two interactions
Eqs.~\eqref{eq:si:umklapp} and \eqref{eq:si:umklapp'}
with the same amount of the layer difference, i.e.,
\begin{equation}
    \tilde{j}-j=\tilde{j}'-j'=\Delta j,
    \label{eq:si:delta_j}
\end{equation}
we get
\begin{equation}
    \Vec{k}^{(\tilde{j}')} 
    = \mathcal{R}((\Delta j) \phi) \Vec{k}^{(j')},
    \label{eq:si:kj'}
\end{equation}
since $\{\Vec{C}_j\}$ and the coordinate vectors
satisfy Eq.~\eqref{eq:(i)} and Eq.~\eqref{eq:newcoordvectors},
respectively.
This indicates that
the surrounding of $\Vec{C}_{\tilde{j}}$,
i.e., the relative direction and distance
from $\Vec{C}_{\tilde{j}}$ to every other $\Vec{k}^{(\tilde{j}')}$
in the momentum-space,
is identical to that of $\Vec{C}_j$,
except the relative rotation of the plane and the coordinate system
by $(\Delta j) \phi$.
And due to the relative rotation of the plane,
the monolayer energies of the states at $\Vec{C}_{\tilde{j}}$ and $\{\Vec{k}^{(\tilde{j}')}\}$
are the same as those at $\Vec{C}_{j}$ and $\{\Vec{k}^{(j')}\}$, respectively.

\begin{figure}[h!]
\centering
\includegraphics[width=0.8\linewidth]{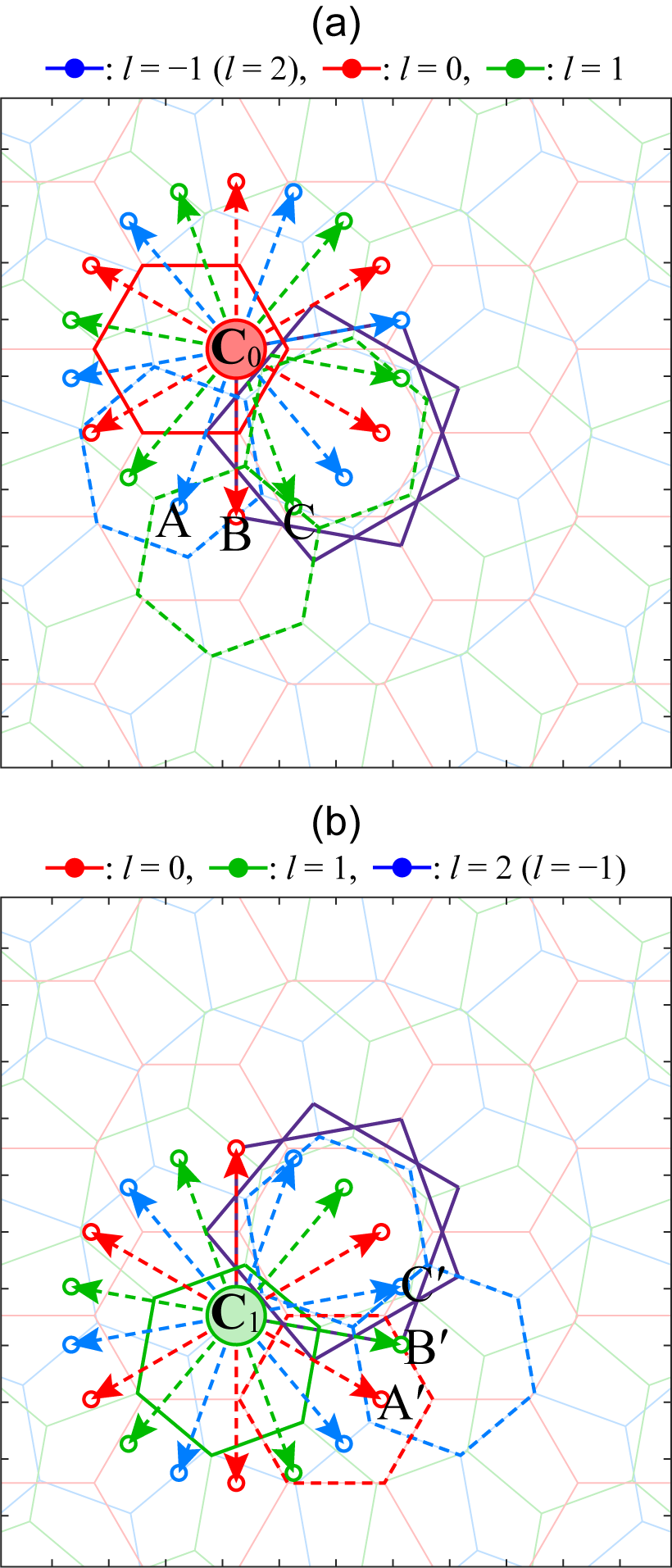}
\caption{
(a) Interlayer interaction from $\Vec{C}_0$ of the resonant chain (purple lines) in Fig.~\ref{Figure_04}(i) in a layer $l=0$.
Here we show only a part of the interactions,
which are associated with the shortest reciprocal lattice vectors of each layer,
among the infinitely many scattering. 
The red, green, and blue hexagons show the extended Brillouin zones of the layer $l=0,1,2$ (=$-1$), respectively. The red dashed arrows show the scattering associated with the reciprocal lattice vectors of the layers $l=0$, which couple the state in $l=0$ to both $l=-1$ and 1 layers. The blue and green arrows show the scattering with the vectors of $l=-1$ and 1, respectively, which couple the state in $l=0$ to $l=-1$ and 1, respectively.
(b) A plot similar to (a) for $\Vec{C}_1$ in a layer $l=1$. The green dashed arrows show the scattering associated with the vectors of $l=1$, which couple the state in $l=1$ to both $l=0$ and 2 layers. The red and blue arrows show the scattering with the vectors of $l=0$ and 2, respectively, and couple the state in $l=1$ to $l=0$ and 2, respectively.
}
\label{Figure_Referee1_Q5_1}
\end{figure}

\begin{figure*}[htb]
\centering
\includegraphics[width=1.0\linewidth]{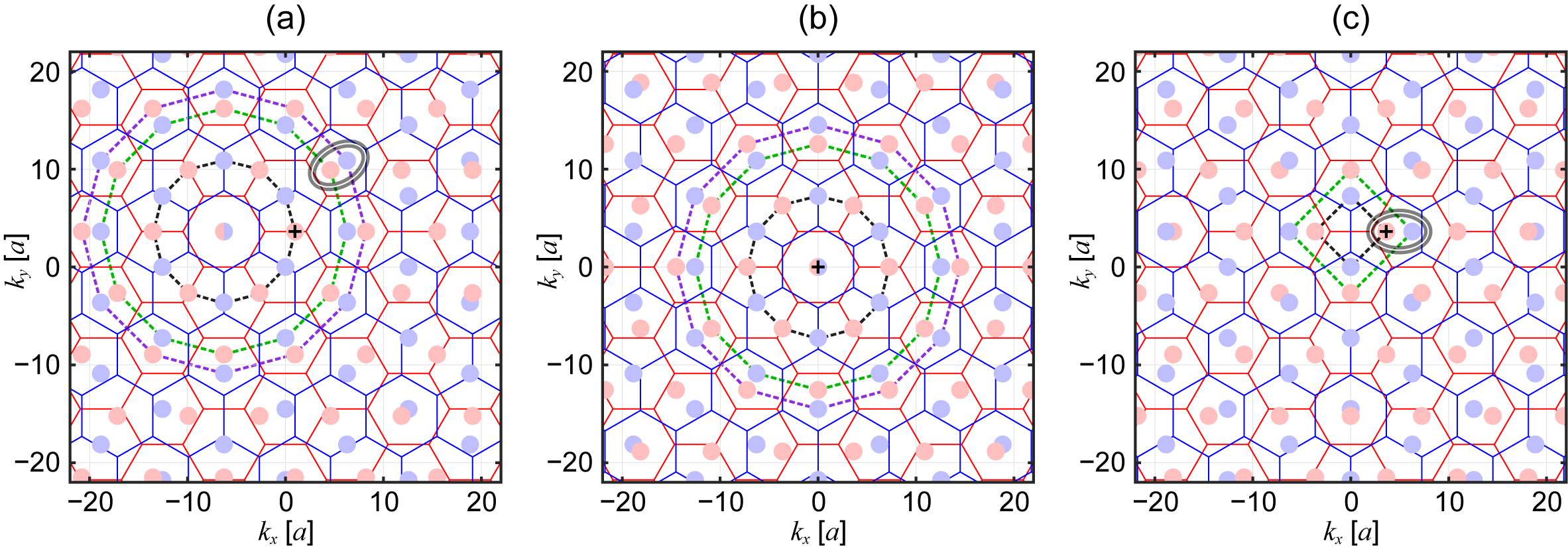}
\caption{
Dual tight-binding lattice in the momentum space for a CTG with $N=2$. The red and blue hexagons show the extended Brillouin zones of the layer $l=0$ and 1, respectively. And the red and blue filled circles represent the states in the layer $l=0$ and 1, respectively, which form the subspace spanned from $\Vec{k}_0$ (cross).
We inverted both the states and the reciprocal lattices of the layer $l=1$ with respect to $\Vec{k}_0$ (see text).
(a) A plot with $\Vec{C}_0$ in Fig.~\ref{Figure_04}(a) as $\Vec{k}_0$,
(b) a plot with $\Vec{k}_0=0$, and
(c) A plot with $\Vec{C}_0$ in Fig.~\ref{Figure_04}(b) as $\Vec{k}_0$.
The black, green, and purple dashed lines in (a) and (b)
show the resonant interaction in Figs.~\ref{Figure_04}(a), \ref{Figure_04}(c), and \ref{Figure_04}(d), 
which have a 12-fold screw rotational symmetry, respectively,
while the black and green dashed lines in (c)
show the resonant interaction in Figs.~\ref{Figure_04}(b) and ~\ref{Figure_04}(e), 
which have a 4-fold screw rotational symmetry, 
respectively.
The gray lines in (a) and (c) show the coupling
between the resonant chains (see text).
}
\label{Figure_Referee1_Q5_2}
\end{figure*}

As an example, we plot a part of the interlayer interactions,
which are associated with the shortest reciprocal lattice vectors of each layer,
from two different $\Vec{C}_j$
of a resonant chain
in Fig.~\ref{Figure_Referee1_Q5_1}.
The red, green, and blue hexagons show the extended Brillouin zones
of the layer $l=0,1,2$ (=$-1$) of a CTG with $N=3$, respectively,
and the purple lines represent the resonant chain
shown in Fig.~\ref{Figure_04}(i)
[$s=1$ and $\Vec{G}^{(0)}=-\Vec{b}_2^{(0)}$].
Figure \ref{Figure_Referee1_Q5_1}(a)
shows the interaction from $\Vec{C}_0$ in a layer with $l=0$.
The red dashed arrows show the scattering associated with the reciprocal lattice vectors of the layers $l=0$, which couple the state in $l=0$ to both $l=-1$ and 1 layers. The blue (green) arrows show the scattering with the vectors of $l=-1$ (1), which couple the state in $l=0$ to $l=-1$ (1).
Figure \ref{Figure_Referee1_Q5_1}(b)
shows the plot similar to (a) for $\Vec{C}_1$
in a layer with $l=1$,
where, the green dashed arrows show the scattering associated with the reciprocal lattice vectors of the layers $l=1$, which couple the state in $l=1$ to both $l=0$ and 2 layers, while the red (blue) arrows show the scattering with the vectors of $l=0$ (2) which couple the state in $l=1$ to $l=0$ (2).
We can first see that the relative position of
$\Vec{C}_0$ to the Brillouin zone of the layer with $l=0$
is the same to that of
$\Vec{C}_1$ to the Brillouin zone of $l=1$,
just as we defined in Sec.~\ref{sec:reson_cond}.
In addition, the relative position of each state
which couples to $\Vec{C}_j$,
i.e., $\Vec{k}^{(j')}$,
to the Brillouin zone
of the corresponding layer is also the same in both figures.
For example, 
$A$ and $C$
in Fig.~\ref{Figure_Referee1_Q5_1}(a)
show the scattering to a state in $l=-1$
and 1 layers, respectively,
while $A'$ and $C'$ in Fig.~\ref{Figure_Referee1_Q5_1}(b)
show the scattering to a state in $l=0$
and 2 layers, respectively.
We can see that the relative position of $A$ ($C$)
to the Brillouin zone of the blue (green) layer
is identical to that of $A'$ ($C'$)
to the Brillouin zone of the red (blue) layer,
as we predicted from
Eqs.~\eqref{eq:si:kj'} and \eqref{eq:newcoordvectors}.
Likewise, $B$ in Fig.~\ref{Figure_Referee1_Q5_1}(a)
shows the scattering to both $l=-1$ and 1 layers,
and $B'$ in Fig.~\ref{Figure_Referee1_Q5_1}(b)
shows the scattering to both $l=0$ and 2 layers.
Again, we can see that
the relative positions of $B$
to the Brillouin zones of the blue and green layers
are identical to those of $B'$
to the Brillouin zones of the red and blue layers, respectively;
note that one of these two scatterings in each figure,
i.e., $B$ to the green layer in
Fig.~\ref{Figure_Referee1_Q5_1}(a)
and $B'$ to the blue layer in
Fig.~\ref{Figure_Referee1_Q5_1}(b),
forms a part of the resonant chain (purple lines).
Thus, each of the states that interacts with $\Vec{C}_j$
has the same monolayer energy as
each of the states that interacts with $\Vec{C}_{\tilde{j}}$.

For $\tilde{j}$ and $\tilde{j}'$ satisfying
Eq.~\eqref{eq:si:delta_j},
Eq.~\eqref{eq:si:matrix_element_of_U'}
becomes
\begin{align}
    &\langle \Vec{k}^{(\tilde{j}')},X'^{(\tilde{j}')} | \mathcal{U} |
    \Vec{C}_{\tilde{j}},X^{(\tilde{j})} \rangle \nonumber\\
    &=-t(\Vec{\tilde{q}}) \, e^{i k_z (j-j') d} \,
    e^{-i\Vec{G}^{(j)}\cdot\mbox{\boldmath \scriptsize $\tau$}_{X}^{(j)}
	+i\Vec{G}^{(j')}\cdot\mbox{\boldmath \scriptsize $\tau$}_{X'}^{(j')}}
\end{align}
by Eq.~\eqref{eq:newcoordvectors}.
This becomes identical to Eq.~\eqref{eq:si:matrix_element_of_U},
if
\begin{equation}
    t(\tilde{\Vec{q}}) = t(\Vec{q}).
    \label{eq:si:tq}
\end{equation}
And Eq.~\eqref{eq:si:tq} is valid for any CTG,
since $t(\Vec{q})$ [Eq.~\eqref{eq_ft}] is isotropic
along the in-plane direction, i.e., $t(\Vec{q})=t(|\Vec{q}|)$,
and
\begin{equation}
    |\tilde{\Vec{q}}|=|\mathcal{R}((\Delta j)\phi) \Vec{q}|
    =|\Vec{q}|.
\end{equation}
As a result, the Hamiltonian matrix element
of each scattering from $\Vec{C}_{\tilde{j}}$ in 
is exactly the same as that from $\Vec{C}_j$.
Thus, the 
surrounding
at $\Vec{C}_{\tilde{j}}$ is identical to that at $\Vec{C}_j$,
and the interaction to the states outside the resonant chain
does not break the degeneracy of the states
forming the chain.
Note that Eq.~\eqref{eq:si:tq} is valid
in any incommensurately stacked atomic layers
with all the dominant interlayer interactions
occur between the atomic orbitals
that have the same magnetic quantum number \cite{PhysRevB.103.045408}.
Thus, 
the resonant states emerge not only in CTG,
but also in most of the Eshelby-twisted atomic layers
composed of transition metal dichalcogenides.

For resonant chains with weak $t(\Vec{q})$,
some interactions to the states outside the resonant chain
can be even stronger than the interaction
forming the chain.
We show some examples of such cases
in Sec.~\ref{sec:offresonance:N=2}.
As shown above, however,
such interactions do not break the symmetry
and degeneracy of the resonant interaction,
and we still get the resonant chains.
Such interactions,
in most cases, 
merely shift the resonant energies
Eqs.~\eqref{eq:E(k0=0):incommensurate}-\eqref{eq:E(k0=0):N=2}
by some, mostly small, constants.

\subsection{CTG with $N=2$}
\label{sec:offresonance:N=2}
While the discussion in Sec.~\ref{sec:offresonance:general}
is valid for any general CTG,
we can obtain further insight on the resonant coupling
by visualizing the resonant chains
in CTGs with $N=2$
in a dual tight-binding lattice.
The subspace spanned by the Hamiltonian $\mathcal{H}$
of a CTG with $N=2$ 
from a wave vector $\Vec{k}_0$ is
$\{|\Vec{k}^{(l)},X^{(l)}\rangle\}$ ($l=0,1$),
where
$\Vec{k}^{(0)}=\Vec{k}_0+\Vec{G}^{(1)}$ and
$\Vec{k}^{(1)}=\Vec{k}_0+\Vec{G}^{(0)}$
[Eq.~\eqref{eq:subspace}].
According to Eq.~\eqref{eq_matrix_element_of_U},
the interaction strength between
$\Vec{k}^{(0)}$ and
$\Vec{k}^{(1)}$ is given by
$t(\Vec{q})$ where $\Vec{q}=\Vec{k}^{(0)}+\Vec{G}^{(0)}=\Vec{k}^{(1)}+\Vec{G}^{(1)}=\Vec{k}^{(0)}-(\Vec{k}_0-\Vec{k}^{(1)})$.
Then, we can visualize the interaction strength
by inverting all the wave points and the extended Brillouin zone
of the $l=1$ layer with respect to $\Vec{k}_0$,
and overlapping them on those of the $l=0$ layer.
Then,
the geometric distance between the wave vectors in the two layers,
$\Vec{k}^{(0)}$ and $\Vec{k}_0-\Vec{k}^{(1)}$,
represents $|\Vec{q}|=|\Vec{k}^{(0)}-(\Vec{k}_0-\Vec{k}^{(1)})|$.
Since $t(\Vec{q})$ decays fast as $|\Vec{q}|$ grows,
the states 
located in a closer distance in the map
exhibit the stronger interlayer interaction.
Thus, if we count each wave vector as a ``site'',
then the arrangement of the wave vectors of the subspace can be
regarded
as a tight-binding lattice in the momentum space,
which is dual to the original Hamiltonian in the real space
\cite{moon2019quasicrystalline}.

Figure \ref{Figure_Referee1_Q5_2}(a) shows
the dual tight-binding lattice of a CTG with $N=2$
by choosing the $\Vec{C}_0$ of the resonant chain
in Fig.~\ref{Figure_04}(a) as $\Vec{k}_0$ (cross mark).
The red and blue hexagons show the extended Brillouin zones of the layer $l=0$ and 1, respectively, and the red and blue filled circles represent $\Vec{k}^{(0)}$ and $\Vec{k}_0-\Vec{k}^{(1)}$,
respectively.
We inverted both the states and the reciprocal lattices of the layer $l=1$ with respect to $\Vec{k}_0$.
The twelve monolayer Bloch states
connected by the black dashed lines are degenerate in energy,
as we can clearly see from the relative positions of the waves
in the Brillouin zone of each layer.
And they interact with the same interaction strength $t(\Vec{q})$,
since they are equally spaced
in the dual tight-binding lattice.
Thus, these states form a resonant chain
as we have already shown in Fig.~\ref{Figure_04}(a).

Each element of the resonant chain interacts 
not only with the elements of the chain
but also with the states outside the chain.
However,
it is obvious that the interaction to the states
outside the chain
neither lifts the degeneracy of the elements
nor breaks the symmetry of the chain.
This is because
the surrounding of each element of the chain is identical,
since all the wave vectors and the extended Brillouin zones
in Fig.~\ref{Figure_Referee1_Q5_2}(a)
are arranged in a 12-fold screw rotationally symmetric way
around the unique center of the dodecagonal lattice,
where the centers of the hexagonal Brillouin zones
of the two layers coincide.
In addition, 
the monolayer state energies of the states outside the chain
are usually different from that of the chain elements.
As we can see from the electronic structures of twisted bilayer graphene
\cite{mele2010commensuration,Moon2013},
if two monolayer Bloch states in different layers have quite different energies,
their interaction gives nearly decoupled, monolayerlike band dispersion
at those wave vectors
no matter how large their $|t(\Vec{q})|$ is.
Thus, the resonant interaction
most strongly influences the electronic structures,
and the interaction to the states outside the chain
merely shifts the resonant energies
Eqs.~\eqref{eq:E(k0=0):incommensurate}-\eqref{eq:E(k0=0):N=2}
by some, mostly small, constants.

%

Note that the chain with the black dashed lines
is not the only resonant chain
that appears
in the map of Fig.~\ref{Figure_Referee1_Q5_2}(a).
There are infinitely many resonant chains
that satisfy the 12-fold screw rotational symmetry
around the center of rotation,
regardless of whether the interaction is strong or weak.
For example, the monolayer Bloch states connected by
the green and purple dashed lines also form
distinct resonant chains.
A careful look at the difference between
the original extended Brillouin zone in Fig.~\ref{Figure_04}
and the inverted Brillouin zone in Fig.~\ref{Figure_Referee1_Q5_2},
we can see that these resonant states are exactly what
we have shown in Figs.~\ref{Figure_04}(c) and (d), respectively.
The distance between the elements in each of these chains
is longer than
that in the chain with the black dashed lines,
which is consistent with the fact that
the chains in Figs.~\ref{Figure_04}(c) and (d) have
weaker $|t(\Vec{q})|$
than the chain in Fig.~\ref{Figure_04}(a).
Interestingly,
these green and purple resonant chains interact rather strongly with each other,
since the distance between the states of the two chains
(encircled by a gray line)
is much shorter
than the distance between the states within each chain.
Thus, the ``coupled sites'' interact with each other
and make up the hybrid resonant chains.
However, since their monolayer state energies are very different,
the coupling merely shifts the resonant energies
by small constants.
Note that there might be a configuration where
the monolayer state forming the resonant chain
is accidentally degenerate with
either state outside the resonant chain.
In that case, the two states form a coupled state,
and the twelve symmetric coupled states form a resonant states.

Also note that the $\Vec{k}_0$ used in Fig.~\ref{Figure_Referee1_Q5_2}(a)
[$\Vec{C}_0$ of the chain in Fig.~\ref{Figure_04}(a)]
is not the only $\Vec{k}_0$ that can reveal
the resonant chains with a 12-fold screw rotational symmetry.
The simplest example is the mapping with $\Vec{k}_0=\Vec{0}$,
which we plot in Fig.~\ref{Figure_Referee1_Q5_2}(b).
In this configuration,
the center of the dodecagonal Brillouin zone lattice appears
at $\Vec{k}^{(0)}=\Vec{k}_0-\Vec{k}^{(1)}=\Vec{0}$,
and all the elements of the subspace are
arranged in a 12-fold screw rotationally symmetric way
around the center.
Accordingly, we get the resonant chains
exactly the same as those in
Fig.~\ref{Figure_Referee1_Q5_2}(a).
In general, 
the subspace spanned by the Hamiltonian $\mathcal{H}$
[Eq.~\eqref{eq:subspace}]
from $\Vec{k}_0'=\Vec{0}+\Vec{G}_0^{(0)}+\Vec{G}_0^{(1)}$
with any $\Vec{G}_0^{(0)} \in \{\Vec{G}^{(0)}\}$
and $\Vec{G}_0^{(1)} \in \{\Vec{G}^{(1)}\}$
is equivalent to that spanned from
$\Vec{k}_0=\Vec{0}$.
This is because
\begin{align}
    \Vec{k}^{(0)}&=\Vec{k}_0+\Vec{G}^{(1)}=\Vec{G}^{(1)}, \nonumber\\
    \Vec{k}^{(1)}&=\Vec{k}_0+\Vec{G}^{(0)}=\Vec{G}^{(0)},
\end{align}
and
\begin{align}
    &\Vec{k}'^{(0)}=\Vec{k}'_0+\Vec{G}^{(1)}=\Vec{G}_0^{(0)}+(\Vec{G}_0^{(1)}+\Vec{G}^{(1)}), \nonumber\\
    &\Vec{k}'^{(1)}=\Vec{k}'_0+\Vec{G}^{(0)}=(\Vec{G}_0^{(0)}+\Vec{G}^{(0)})+\Vec{G}_0^{(1)},
    \end{align}
and the sets of the monolayer Bloch states
$\{|\Vec{k}'^{(0)},X^{(0)}\rangle\}$
and $\{|\Vec{k}'^{(1)},X^{(0)}\rangle\}$
are equivalent to
$\{|\Vec{k}^{(0)},X^{(0)}\rangle\}$ and
$\{|\Vec{k}^{(1)},X^{(0)}\rangle\}$,
up to the Bloch phase,
respectively.
Likewise,
$\{\Vec{k}'^{(0)}\}$ and $\{\Vec{k}'_0-\Vec{k}'^{(1)}\}$
form a dual tight-binding lattice
which is equivalent to the lattice composed of
$\{\Vec{k}^{(0)}\}$ and $\{\Vec{k}_0-\Vec{k}^{(1)}\}$,
except a finite shift of the entire lattice
along the in-plane direction;
the former lattice is shifted by $\Vec{G}_0^{(0)}$
from the latter lattice.
For example, since the $\Vec{k}_0$ used in
Fig.~\ref{Figure_Referee1_Q5_2}(a)
is the $\Vec{C}_0$ in Fig.~\ref{Figure_04}(a),
and $\Vec{C}_0=-\Vec{b}_1^{(0)}+\Vec{b}_1^{(1)}$ therein,
the lattice in Fig.~\ref{Figure_Referee1_Q5_2}(a)
is merely shifted from
that in Fig.~\ref{Figure_Referee1_Q5_2}(b)
by $-\Vec{b}_1^{(0)}$.

Although the infinitely many combinations of the incommensurate
$\Vec{G}_0^{(0)}$ and $\Vec{G}_0^{(1)}$
give infinitely many $\Vec{k}_0'$,
such $\{\Vec{k}_0'\}$ do not cover
every point in the Brillouin zone.
This is because the cardinality of the points
in the Brillouin zone is $\aleph_1$,
while that of such $\{\Vec{k}_0'\}$ is $\aleph_0$,
and $\aleph_0 < \aleph_1$.
Thus, there are other interactions
which are not shown in Figs.~\ref{Figure_Referee1_Q5_2}(a) and (b).
Figure \ref{Figure_Referee1_Q5_2}(c)
shows one of such examples.
Here, we plot the dual tight-binding lattice
by choosing the $\Vec{C}_0$ of the resonant chain
in Fig.~\ref{Figure_04}(b),
$\Vec{C}_0=\Vec{b}_2^{(0)}/2+\Vec{b}_1^{(1)}/2$,
as $\Vec{k}_0$.
The lattice
does not host a dodecagonal center,
since $\Vec{b}_2^{(0)}/2 \not \in \{\Vec{G}^{(0)}\}$
and $\Vec{b}_2^{(1)}/2 \not \in \{\Vec{G}^{(1)}\}$;
instead, it exhibits a 4-fold screw rotational symmetry.
We plot the first two strongest resonant interactions
having a 4-fold screw rotational symmetry,
which correspond to the resonant chains in
Figs.~\ref{Figure_04}(b) and (e),
by black and green lines, respectively.
Again, these two chains also interact with each other
(gray line),
but such a coupling merely shifts the resonant energies,
since their monolayer state energies are very different.



\section{Eigenvectors of the resonant states}
\label{sec:c1_and_c2}

The $c_{m,1}^\pm$ and $c_{m,2}^\pm$ of Eq.~\eqref{eq:psi(k0=0)} are
\begin{equation}
    \begin{pmatrix}
    c_{m,1}^\pm \\ c_{m,2}^\pm
    \end{pmatrix}
    =
    \begin{pmatrix}
    c_a \\
    -c_b \pm \sqrt{|c_a|^2 + c_b^2}
    \end{pmatrix}/c,
\end{equation}
where
\begin{align}
    c_a &= h_0 -2t_0 \tilde{\omega} \cos q_m, \nonumber\\
    c_b &= 2t_0 \sin q_m \sin q_G,
    \label{eq:c1_and_c2:supp1}
\end{align}
for incommensurate CTGs,
\begin{align}
    c_a &= h_0 -2t_0 \tilde{\omega} \cos (q_z+q_m), \nonumber\\
    c_b &= 2t_0 \sin (q_z+q_m) \sin q_G,
    \label{eq:c1_and_c2:supp2}
\end{align}
for commensurate CTGs with $N \ne 2$, and
\begin{align}
    c_a &= h_0 -4t_0 \tilde{\omega} \cos q_z \cos q_m, \nonumber\\
    c_b &= 4t_0 \cos q_z \sin q_m \sin q_G,
    \label{eq:c1_and_c2:supp3}
\end{align}
for commensurate CTGs with $N=2$,
and $c$ is the normalization constant.
The parameters in Eqs.~\eqref{eq:c1_and_c2:supp1}-\eqref{eq:c1_and_c2:supp3} are
defined in Sec.~\ref{sec:bs_and_psi}.



\bibliography{qc}

\end{document}